\documentclass[final]{siamltex}

\usepackage[]{fontenc}
\usepackage[latin1]{inputenc}
\usepackage{verbatim}
\usepackage{amsmath}
\usepackage{amssymb}
\usepackage{dsfont}
\usepackage{amsfonts}
\usepackage{mathrsfs}
\usepackage{graphicx}
\usepackage{color}
\usepackage{ulem}
\usepackage{xspace,bbm,epic,eepic}
\usepackage{pgfplots}
\usepackage{setspace}
\usepackage{xspace}

\usepackage{ stmaryrd }
\usepackage{graphicx}
\usepackage{lineno}
\usepackage{color}
\usepackage{url}
\usepackage{graphicx}
\usepackage{ stmaryrd }
\usepackage{algorithm}
\usepackage{algpseudocode}

\usepackage{amsmath,amssymb,color} 

\newcommand{\Rr}{\mathds{R}}

\newcommand{\unit}{1\!\!1}

\newcommand{\eg}{\textit{e.g.}, }
\newcommand{\ie}{\textit{i.e.}, }

\DeclareMathOperator*{\argmax}{arg\,max}
\DeclareMathOperator*{\argmin}{arg\,min}

\newtheorem{remark}{Remark}

\def\Spwidth{{\epsilon}}		

\def\dimSps{{K}} 		        	
\def\Sps{{V}}				

\def\idxss{{k}}
\title{ Selecting Reduced Models\\ in the Cross-Entropy Method}

\author{P. H\'eas
\thanks{INRIA \& IRMAR,  Univ Rennes, Campus de Beaulieu, 35042 Rennes, France. ({\tt patrick.heas@inria.fr}). 
} 
   }
\date{}
\begin{document}
%
\maketitle
\begin{abstract}
This paper deals with the estimation of rare event probabilities using importance sampling (IS), where an {\it optimal} proposal distribution is computed with the cross-entropy (CE) method. Although, IS optimized with the CE method leads to an efficient reduction of the estimator  variance, this approach  remains  unaffordable for problems where the repeated evaluation of the score function represents a too intensive computational effort.   This is often the case  for score functions related to the solution of a partial differential equation (PDE)  with random inputs. 
This work proposes to alleviate  computation  by the parsimonious use of a hierarchy of score function approximations  in the CE optimization process. The score function approximation is obtained by selecting the surrogate of lowest dimensionality, whose accuracy guarantees to pass the  current CE optimization stage.  The selection  relies on certified upper bounds on the  error norm. An asymptotic analysis provides some theoretical guarantees on the efficiency and convergence of the proposed algorithm. Numerical results demonstrate the gain brought by the  method in the context of pollution alerts and a system modeled by a PDE.

\end{abstract}
\begin{keywords}
Rare event simulation, importance sampling, cross-entropy optimization, reduced basis, certified error bound, partial differential equation.%
\end{keywords}

\section{Introduction}
The accurate estimation of the probability of a rare event  with standard Monte Carlo typically requires the evaluation of a score function for a very large set of points: the number of points is of the order of the inverse of the sought  rare event probability. The evaluation of the score function for this very large set  becomes particularly infeasible if for each point a computationally expensive model is involved. 

IS
is a variance reduction strategy for Monte Carlo estimation. The idea is to sample from a  biasing distribution, such that fewer samples
are necessary to obtain the same accuracy of the rare event probability than with standard
Monte Carlo. The bias introduced by the sampling from the biasing distribution is corrected
by reweighing the samples in the IS estimator. 

The key of the performance of IS  lies in the choice of  the problem-dependent so-called {\it biasing} or {\it proposal} distribution. Although the optimal biasing distribution that leads to a zero variance IS  estimator is closed-form,  computing this zero-variance density is not straightforward since
it requires the sought probability of the  rare event. The
CE method \cite{rubinstein1997optimization,Homem-de-Mello02rareevent,Rubinstein:2007:SMC:1349778,de2005tutorial} provides an efficient way to approximate this zero-variance density within a parametric family of probability distributions. The CE method searches the {\it optimal}\, biasing distribution, in the sense that it will minimise the Kullback-Leibler
divergence from the zero-variance density among the feasible distributions. Even though, computing the biasing distribution with the CE method may still be prohibitive if the  model involved in the score function is computationally demanding.
 

This paper is concerned by the following question:   can we save  computational power by using score function approximations in the CE method?  
Several works  recommend score function approximations in order to accelerate the estimation of rare event probabilities. In particular, score function approximations have already been put forward in IS~\cite{bichon2011efficient,dubourg2013metamodel,PEHERSTORFER2016490, peherstorfer2015online,wu2016surrogate}.\footnote{We note that 
such approximations have also been studied in the context of multilevel splitting~\cite{papadopoulos2012accelerated,ullmann2015multilevel}. However, an exhaustive discussion of all possible  combinations of alternative estimation techniques with score function approximations is  out of the scope of this work.} Nevertheless, to the best of our knowledge, the methods proposed in \cite{peherstorfer2017multifidelity} and \cite{LI20118683} are the unique works directly concerned by the raised question and providing partial  answers. 
 The authors in \cite{peherstorfer2017multifidelity} define a sequence of  auxiliary ``low-fidelity  CE methods''  using a set of score function approximations. The auxiliary low-fidelity CE methods are then run  sequentially in order to  pre-condition the standard CE method defined with the original score function.  Under certain conditions on  score function  approximations, the pre-conditioning  guarantees that the  number of iterations  is lowered at each of the auxiliary CE method levels, or at worst remains identical. Nevertheless, using this pre-conditioner does not necessarily guarantee a reduction of the global computational cost for computing the biasing distribution.  The pre-conditioner is in fact  an initialization strategy rather than a way of integrating  and adjusting the score function approximations  in the CE method. Computational power may still be spoilt in the case an accurate estimate is not needed at some levels of   the  pre-conditioning sequence.  
  In \cite{LI20118683}, the authors propose an IS estimator based on  an hybrid ``low-fidelity / high-fidelity  CE method''. The idea  is to approximate the score function, by dividing the probability space into two sub-domains. Regions which are close in some sense to the rare event are evaluated with the high-fidelity model. The remaining part of the probability space is evaluated using a low-cost surrogate model. However, parameters of the proposed algorithms are exposed to arbitrariness,
 leading to non-certified sub-domain discrimination, which in turns implies the computation of a sub-optimal biasing distribution. 

In this work, we propose a CE algorithm based on score function approximations  which converges almost surely to the optimal  biasing distribution.  A surrogate is selected in a hierarchy of precomputed score function approximations at each iteration of the CE method, depending on the needed accuracy. As for the certified  reduced-basis evaluation of failure probability in \cite{Chen:196807}, the  selection of the score function approximations exploits  upper bounds on the error norm of the reduced model. We focus  on {\it reduced basis} (RB) approximations, for which  there exists  an extensive literature on the computation of  {\it a priori} or {\it a posteriori} error estimates for approximation of the solution of a PDE using RB, see \eg \cite{quarteroni2015reduced}. Besides, we also  provide an asymptotic analysis showing  that under mild conditions {\it i)}   for each of the algorithm iterations,     the squared coefficient of variation (SCV)  of the IS estimator  is  the minimal achievable with the current score function approximation, {\it ii)} the convergence is guaranteed towards the optimal biasing distribution  in at most the same number of  iterations as the standard  CE method. 

The paper is organized as follows. Section~\ref{sec:ingredients} recalls  the basics of  IS for rare event estimation with the CE method. It then introduces score function approximations and reduced models. In Section~\ref{sec:recipes}, we present state-of-the-art approaches, and in particular the method pre-conditioning the CE method. We then propose in Section~\ref{sec:ourRecipes} our  algorithm. A theoretical result attesting of the asymptotic performance of our  method is given in Section~\ref{sec:theorie}. We provide details on the proof  of  the proposed theorem in the appendices. 
Section~\ref{sec:numerical} presents the numerical evaluation in the case of rare event probability estimation related to a pollution alert problem.  We finally provide  concluding remarks in a last section. 

\section{Efficient Rare Event Estimation by Importance Sampling}\label{sec:ingredients}
 We assume that $X$ is a random element taking its values in $\Rr^p$ and denote by $\mu$ its probability distribution.
 We denote by $A$ the  set of rare events of interest, and we assume that $$A=\{x\in \Rr^p : \phi(x) \ge \gamma^\star\},$$ for some real number $\gamma^\star$ and for a score function $\phi: \Rr^p \to \Rr$. 
The probability of the rare event $ X \in A$ is defined as 
$
p_A=\langle \unit_{A}, \mu \rangle, 
$
where $\langle \cdot , \mu \rangle $ denotes an integration with respect to the probability measure $\mu$.
 We look for an estimator of $p_A$ where $\gamma^\star$ is large so that $p_A \ll 1$. We  assume that we know how to draw $m$ independent and identically distributed  (i.i.d.) samples  $x_1,\cdots,x_m$  from $\mu$.
 \subsection{First Ingredient: the Cross-Entropy  Method}
  The naive Monte-Carlo (MC) estimator of the rare event probability is
\begin{align}\label{eq:naiveMC}
 p_A^{MC}=\frac{1}{m}\sum_{i=1}^m \unit_{A}(x_i).
\end{align}
This estimator  is unbiased. Its relative error is measured by the SCV of the estimator, which is 
\begin{align*}
\frac{\mathbb{E}[( p_A^{MC}-p_A)^2]}{ p_A^2}=\frac{1}{m}\left(\frac{1}{p_A}-1\right)
&\overset{p_A \ll 1}{\simeq} \frac{1}{{mp_A}}
\end{align*}
so that we need $m > p_A^{-1}$ for a relative error smaller than one.

Let $\textrm{supp}(\mu)=\{x\in \Rr^p : \mu(x)>0\}$ be the support of the distribution $\mu$. For a biasing distribution $\nu$ with $\textrm{supp}(\mu) \subseteq  \textrm{supp}(\nu)$, the IS estimator  $  p^{IS}_{A, \nu}$ of  $p_A$ is 
\begin{align}\label{eq:IS}
  p^{IS}_{A,\nu}=\frac{1}{m}\sum_{i=1}^m  \unit_{A}(z_i)\frac{\mu(z_i)}{\nu(z_i)},
\end{align}
with $m$  i.i.d.  samples $z_1,\cdots,z_m$ from $\nu$. It is easy to see that $  p^{IS}_{A,\nu}$ is an unbiased  estimator of $p_A$. Its SCV is 
\begin{align}\label{eq:SCV}
\frac{\mathbb{E}[( p_{A,\nu}^{IS}-p_A)^2]}{ p_A^2}=\frac{1}{mp_A^2}\textrm{var}( \unit_{A}\frac{\mu}{\nu}, \nu).
\end{align}
The optimal biasing distribution $\nu_A^\star$ yielding a zero-variance estimator (\ie a zero SCV)  is 
\begin{align}\label{eq:optDensity}
\nu_A^\star= \frac{\unit_{A}\mu}{p_A}.
\end{align}
Unfortunately, it depends  on the rare event probability $p_A$ we want to estimate.
Now, consider a set of parametrized densities $\mathcal{V}= \{\nu^\theta : \theta \in  \Theta \}.$
The CE method~\cite{rubinstein1997optimization} optimizes for a parameter $\theta^\star \in \Theta$ such that the corresponding distribution $\nu_A^{\theta^\star}$  minimizes the Kullback-Leibler divergence from the zero-variance density~$\nu_A^\star$:
\begin{align}\label{eq:minDiv}
\nu_A^{\theta^\star} \in \argmin_{\nu^\theta \in \mathcal{V}} \langle  \ln \frac{\nu_A^\star}{\nu^\theta}, \nu_A^\star\rangle.
\end{align}
In the desirable situation where $\nu^\star_A \in \mathcal{V}$, a minimizer $\theta^\star$ of~\eqref{eq:minDiv} is such that $\nu^{\theta^\star}_A=\nu^\star_A$. Using~\eqref{eq:optDensity}, the optimization problem~\eqref{eq:minDiv} can be rewritten as 
\begin{align}\label{eq:minDiv2}
\nu_A^{\theta^\star} \in \argmax_{\nu^\theta \in \mathcal{V}} \langle  \unit_{A} \ln \nu^\theta, \mu \rangle.
\end{align}
Solving the stochastic counterpart of~\eqref{eq:minDiv2} 
\begin{align}\label{eq:minDiv3}
\hat\nu_A^{\theta^\star} \in  \argmax_{\nu^\theta \in \mathcal{V}} \frac{1}{m} \sum_{i=1}^m \unit_{A}(x_i) \ln \nu^\theta(x_i),
\end{align}
with $m$  i.i.d.  samples $x_1,\cdots,x_m$ from $\mu$, typically fails because~\eqref{eq:minDiv3} is affected by the rareness of the event $\unit_{A}(X)=1$, just as for the naive MC estimator~\eqref{eq:naiveMC}.
To circumvent this effect, starting from the initial distribution $\nu_0=\mu$ and for some parameter $\rho \in (0,1)$, the CE method estimates sequentially  a sequence of nested events \vspace{-0.3cm}
\begin{align}\label{eq:nestingOrig}
A_1 \supseteq A_2 \supseteq  \cdots \supseteq A,
\end{align}
 such that  $A_{j+1}=\{x\in \Rr^p : \phi(x) \ge  \gamma(\nu_j,\rho, \phi) \}$ with the $(1- \rho)$-quantile 
\begin{align*}
 \gamma(\nu_j,\rho, \phi)&=\max\{s\in \Rr : \langle \unit_{\phi(\cdot) \ge s}, \nu_j \rangle \ge \rho \}=\max\{s\in \Rr : \langle \unit_{\phi(\cdot) < s}, \nu_j \rangle \le 1- \rho \} ,
\end{align*}
 and jointly updates the biasing distribution $\nu_j$ according to 
  \begin{align}\label{eq:LoptimProposDist_j}
\nu_j \in  \argmax_{\nu^\theta \in \mathcal{V}}\langle \unit_{A_{j}} \frac{\mu}{\nu_{j-1}} \ln \nu^\theta, \nu_{j-1} \rangle.
\end{align}
In words, the occurrence of the event $\unit_{A_j}(X)=1$ where $X$ is a random variable of distribution $\nu_{j-1}$ tends on the one hand to decrease since we have $A_{j-1}\supseteq A_{j}$. On the other hand, it tends to increase since,  according to~\eqref{eq:nestingOrig} -~\eqref{eq:LoptimProposDist_j}, $\nu_{j}$ is  nearer (in terms of cross-entropy) from $\nu_A^\star$ than $\nu_{j-1}$. Typically, a proper setting for this tradeoff will yield an occurrence of the event $\unit_{A_j}(X)=1$ that varies little with the algorithm iterations. Then, if the set $A_1$ and the initial distribution are chosen so that  the event $\unit_{A_1}(X)=1$ is not rare, we can typically expect that the solution of the stochastic counterpart of problem~\eqref{eq:LoptimProposDist_j} will be a ``good'' approximation of $\nu_j$, and that the CE method will yield a ``good'' approximation of $\nu_A^{\theta^\star} $.

\subsection{Second Ingredient: Score Function Approximations}\label{sec:2ndIng}

 We consider a {\it high-fidelity} model $f^\star:\Rr^p \to \Rr^q$ parametrized by  $ x   \in \Rr^p$.  Let  $\mathcal{X} \subset \Rr^p$ be the parameter space. Furthermore, we assume that we have at our disposal  a set of model evaluations $\{f^\star( x_i )\}_{i=1}^N$, so-called {\it snapshots}, corresponding  to $N$ representative samples $x_i$ of the parameter space $\mathcal{X}$.
We focus in this work on  reduced modeling  strategies, which uses the set of snapshots to construct an approximation subspace $\Sps_\dimSps$ for the set 
$
\{f^\star( x ) :  x  \in \mathcal{X}  \}.
$ 
Typical examples of such strategies are the common  greedy RB methods or principal orthogonal decomposition (POD).
Those standard  techniques usually generate  the approximation subspace $\Sps_\dimSps$  together with a sequence of nested RB spaces 
$
\Sps_1\subset \Sps_2\subset \ldots \subset \Sps_\dimSps, 
$
 of increasing dimension $d_k=\dim(V_{k})$.
A set $\{ f^{(k)}(x_i)\}^{K}_{k=1}$ of low-dimensional approximations  of the snapshot $f^\star(x_i)$, also called reduced models or surrogates, is then computable using the sequence of nested subspaces. For instance,  $f^\star( x_i )$ can be a finite element discretization  of a  PDE and   $ f^{(k)}(x_i)$ some {\it Petrov-Galerkin}  or  {\it least-square} RB approximation~\cite{quarteroni2015reduced}.  
We will adopt the following general notations: we assume available a hierarchy $\mathcal{F}=\{ f^{(1)}, f^{(2)},\ldots, f^{(K)}, f^{(K+1)}\}$  of  models  defined over $\Rr^p$ and taking their values in $\Rr^q$, the high-fidelity model $f^{(K+1)}=f^\star$ being the last element of the set. The hierarchy of model is such that for $k=1,\ldots, K$,   we have   $d_k<d_{k+1}\in \mathcal{K}$, with $d_{K+1}=q$ and $\mathcal{K}$ denoting a prescribed set of ordered positive integers lower or equal to $q$.

From the computational standpoint, in the case $f^\star( x )$ is a finite element discretization  of an elleptic PDE with $r$ affine  dependences in the bilinear  form parametrized by $x$,  computing a  Petrov-Galerkin reduced-basis approximation  $ f^{(k)}$  at point $x$ requires  off-line $\mathcal{O}(r d_k q^2)$ operations and, more importantly, it requires on-line  $\mathcal{O}(d_k^2(r+d_k))$ operations. Computing the high-fidelity solution requires instead $\mathcal{O}(q^3)$ operations, which suggests that an important  computation gain for RB approximation if $d_k \ll q$. We remark  that the complexity is slightly higher for a least-square reduced-basis approximation. The possibility to devise an off-line/on-line decomposition relies on the assumption of affine parametric dependence. Nevertheless, similar low on-line complexities are achievable in the case of non-affine parametric dependence using approximations by means of the {\it empirical interpolation method}. We refer to~\cite{quarteroni2015reduced} for details and further extensions to the case of non-linear PDEs with non-affine parametric dependences.

Finally, we assume  a score function $\phi: \Rr^p \to \Rr$  of the form $\phi( x )=h(f^\star( x ))$, where  $h:\Rr^q \to \Rr$. Using the reduced models, we  define score function approximations   $\phi^{(k)}: \Rr^p \to \Rr$ as $\phi^{(k)}( x )=h(f^{(k)}( x ))$. Therefore,  the hierarchy $\mathcal{F}$ is related to a sequence of approximations $\mathcal{M}=\{ \phi^{(1)}, \phi^{(2)},\ldots, \phi^{(K)}, \phi^{(K+1)}\},$  where the $(K+1)$-th element is the original score function  $\phi^{(K+1)}=\phi$.

\subsection{Third Ingredient: Error Bounds}
For any    $x  \in \mathcal{X}$, we further assume  that we can compute  on-line for $k=1,\ldots,K$ an error bound of the form
\begin{align}\label{eq:aPosteriori}
\| f^\star( x ) -  f^{(k)}( x )\|_2 \le  \Spwidth_\idxss(x),
\end{align} 
with a complexity independent of the high-fidelity model dimension.
Although the evaluation of error bounds can be in some cases expensive,  error estimators  for RB approximation involves generally low-complexity computations and are largely  employed to steer adaptive schemes, see \eg \cite{quarteroni2015reduced} and references therein.    Indeed,
taking advantage of the  previously introduced off-line/on-line decomposition, it is typically possible to compute efficiently for any $x \in \mathcal{X}$ a local error bound for each subspace approximation.  {\it A posteriori error estimates}  \cite{quarteroni2015reduced} are computable on-line with a quadratic complexity in $d_k$ and $r$. More precisely, the   error norm of the RB approximation is  bounded by the ratio of the dual norm of the residual  and  the stability factor. The dual norm of the residual  can  typically be evaluated on-line with $\mathcal{O}(r^2d_k^2)$ operations, exploiting affine~\cite[Chapter 3.7]{quarteroni2015reduced} or approximate affine~\cite[Chapter 10.4]{quarteroni2015reduced} parametric dependance. On the other hand, provided an off-line characterization by {\it interpolatory  radial basis functions}, the stability factor term  is evaluated on-line with a complexity linear in the number of interpolation points and independent of $q$.

Now, in the case $h$ is a bounded linear function, by definition of an operator  norm, we have 
$
| \phi( x ) -\phi^{(k)}( x ) | \le \| h \|_{2,2} \|  f^\star( x ) -  f^{(k)}( x )\|_2.
$
In the case $h$ is the sup norm,  using the triangular inequality and the fact that  $f^\star( x )\in \Rr^q$ with $q$ finite, we obtain
$| \phi( x ) -\phi^{(k)}( x ) | \le \|  f^\star( x ) -  f^{(k)}( x )\|_\infty \le \|  f^\star( x ) -  f^{(k)}( x )\|_2.$
Using the  error estimate~\eqref{eq:aPosteriori}, we can then bound the error on any subset $\mathcal{Z}$ of $\mathcal{X}$:\vspace{-0.cm}   
\begin{align}\label{alpha_k}
\forall\, x \in \mathcal{Z}, \quad |\phi( x ) -\phi^{(k)}( x )| \le \alpha_k( \mathcal{Z})= c \max_{x'\in \mathcal{Z}}\epsilon_k(x').\vspace{-0.2cm}
\end{align}
where $c=\| h \|_{2,2}$ or $c=1$ respectively for linear operators or for the sup norm. For ease of notations, when dealing with a distribution $\nu \in \mathcal{V}$, we will use  the simplified  notation $\alpha_{k}(\nu)$, instead of  $\alpha_{k}(\supp(\nu))$.

We mention that the {\it goal-oriented} adaptive strategy proposed in~\cite{Chen:196807}  refines the error bound estimate~\eqref{alpha_k}. The idea of such a strategy  is to directly rely on a posteriori estimates of the score function approximation error in order to infer a sequence of subspaces yielding  fine approximations of the score function close to the  set of rare events of interest and coarse approximations far away from it.

\section{State-of-the-Art Recipes}\label{sec:recipes}
\begin{algorithm}[H]
\begin{algorithmic}[0]
\State \textbf{input parameters}: ($m$, $\rho$, $\delta$, $\nu_0$)

 \State {{\it \bf 1}- Draw $m$  i.i.d.  samples $z_0,\cdots,z_{m}$ from $\nu_0$.}
\State {{\it \bf 2}- Compute  $\phi(z_i)$'s.} 
\State {{\it \bf 3}- Set   $j=0$, $m_0=m$ and $\rho_0=\rho$.}
 \While{
 $\gamma( \hat \nu_{j},\rho_{j},\phi) < \gamma^{\star}$
 }
\State{{\it \bf 3.1}- Set  $ j=j+1$ and solve\vspace{-0.5cm}
 \begin{align}\label{eq:minDiv4}
&\nu_j \in  \argmax_{\nu^\theta \in \mathcal{V}} \frac{1}{m_{j-1}} \sum_{i=1}^{m_{j-1}} \unit_{\hat A_j}(z_i) \frac{\mu(z_i)}{\nu_{j-1}(z_i)} \ln \nu^\theta(z_i).
\end{align}}
\State {{\it \bf 3.2}- Draw $m_{j-1}$  i.i.d.  samples $z_1,\cdots,z_{m_{j-1}}$ from $\nu_j$.}
\State {{\it \bf 3.3}- Compute  $\phi(z_i)$'s.}
\State {{\it \bf 3.4}-   Adapt  $m_j$ and  $\rho_j$ starting from  $m_j=m_{j-1}$ and $\rho_j=\rho_{j-1}$   such that \begin{align}\label{eq:condCEstandard}\gamma(\hat \nu_j, \rho_j , \phi) \ge \bar \gamma= \min\{\gamma^\star, \gamma(\hat \nu_{j-1},\rho_{j-1}, \phi )+\delta\},\end{align} \hspace{1.2cm} using Algorithm~\ref{algo:1part1}.
  }

 \EndWhile
 \State{{\it \bf 4}- Set  $ J=j+1$ and solve~\eqref{eq:minDiv4} with $\hat A_J=A$}.

\State{{\it \bf 5}-  Draw $m_{J-1}$  i.i.d.  samples $z_1,\cdots,z_{m_{J-1}}$ from $\nu_J$.}
\State {{\it \bf 6}- Compute  $\phi(z_i)$'s.}
\State {{\it \bf 7}- Set $\hat\nu_{A}^{\theta^\star}=\nu_J$ and compute $  p^{IS}_{A,\hat\nu_{A}^{\theta^\star}}$ using~\eqref{eq:IS}  for $\nu=\hat\nu_{A}^{\theta^\star}$}.
\State \textbf{outputs}:  ($\hat\nu_{A}^{\theta^\star}$, $  p^{IS}_{A,\hat\nu_{A}^{\theta^\star}}$).
\end{algorithmic}
\caption{ - Standard CE Algorithm \cite{Homem-de-Mello02rareevent} \label{algo:1}}
\end{algorithm}

\begin{algorithm}[H]
\begin{algorithmic}[0]
\State \textbf{inputs}: ($m_{j}$, $\rho_{j}$, $z_i$'s,  $\beta$)

\While{\eqref{eq:condCEstandard} does not hold}
\If{ $\exists \bar \rho $  satisfying  $\gamma(\hat \nu_j, \bar \rho ,  \phi) \ge \bar \gamma$}
 \State {{\it \bf 1}- Set $\rho_j$ as the largest existing $\bar \rho$.}
\Else
 \State {{\it \bf 2}- Draw $\lceil \beta m_j \rceil -m_j$ additional  i.i.d.  samples $z_{m_j+1},\cdots,z_{\lceil \beta m_j \rceil}$ from $\nu_j$.}
\State {{\it \bf 3}-  Set $m_j=\lceil \beta m_j \rceil $. } 
\State {{\it \bf 4}-  Compute  $\phi(z_i)$'s. } 
\EndIf
\EndWhile
\State \textbf{ouputs}: ($m_{j}$, $\rho_{j}$, $z_i$'s)


\end{algorithmic}
\caption{ -  Adaptation of  $m_j$ and $\rho_j$ \cite{Homem-de-Mello02rareevent}  \label{algo:1part1}}
\end{algorithm}

\begin{algorithm}[H]
\begin{algorithmic}[0]
\State \textbf{input parameters}: ($m$, $\rho$, $\delta$, $\nu^{(0)}$, $\mathcal{M}$)
\For{level $k=1,\cdots,K+1$}
\State{{\it \bf 1.1}-  Substitute $\phi$ by $\phi^{(k)}$ in Algorithm~\ref{algo:1}}. 
\State {{\it \bf 1.2}- Compute  $\nu^k={\hat\nu_{A}^{\theta^\star}}$ with  Algorithm~\ref{algo:1} initialized with $\nu_0=\nu^{k-1}$}.
\EndFor
\State{{\it \bf 2}- Compute $  p^{IS}_{A,{\nu_{A}^{\theta^\star}}}$  using~\eqref{eq:IS}  for $\nu=\nu^{K+1}$}.
\State \textbf{outputs}:  ($\hat\nu_{A}^{\theta^\star}$, $  p^{IS}_{A,\hat\nu_{A}^{\theta^\star}}$).
\end{algorithmic}
\caption{ - Pre-conditioned    CE algorithm \cite{peherstorfer2017multifidelity} \label{algo:3}}
\end{algorithm}

Incorporating one after the other the two first ingredients in IS, we obtain the standard and the  pre-conditioned CE algorithms. Note that we review hereafter  algorithms which are certified to converge to the optimal biasing distribution. In consequence, we do not detail the hybrid  method suggested in~\cite{LI20118683},   although it relies on the same ingredients and is related, up to some extent.

\subsection{Standard CE Algorithm}

The standard CE method is exposed  in Algorithm~\ref{algo:1}.   The algorithm builds a sequence of proposal distributions starting from an initial distribution $\nu_0$ and ending with the optimal proposal $\hat\nu_{A}^{\theta^\star}$. 

We detail hereafter this construction.   In the first steps of the algorithm, $m$ samples are drawn from the initial proposal $\nu_0$ (in step 1) and related scores are computed (in step 2). Then, the algorithm estimates (in step 3) a sequence of proposals $\{\nu_1,\ldots, \nu_{J-1}\}$, and the related sequences of  quantile parameters $\{\rho_1,\ldots, \rho_{J-1}\}$ and sample sizes $\{m_1,\ldots, m_{J-1}\}$,  until the quantile reaches the desired score $\gamma^{\star}$.
More precisely, at the $j$-th iteration, given $\rho_j \in (0,1)$, a sample size $m_j$ and a proposal $\nu_j$, the algorithm builds the empirical $(1-\rho_j)$-quantile \vspace{-0.cm}
\begin{align}\label{eq:quantile}
  \gamma(\hat \nu_j, \rho_j, \phi ) =\max\{s\in \Rr : \left( \frac{1}{m_j} \sum_{i=1}^{m_j} \unit_{\phi(z_i) < s}  \right) \le  1-\rho_j \},
 \end{align}
 where  $z_1,\cdots,z_{m_j}$ are    i.i.d.  samples from $\nu_j$. This quantile  defines the stochastic event \vspace{-0.25cm}
  \begin{align}\label{eq:setAj}
  \hat A_{j+1}&=  \{x\in \Rr^p : \phi(x) \ge \min\left(  \gamma(\hat \nu_{j},\rho_{j}, \phi) , \gamma^\star \right) \}.
\end{align}
Given  the set $\hat A_{j+1}$ and the proposal  $\nu_{j}$,  a refined distribution $\nu_{j+1}$ is estimated (in step 3.1) by solving~\eqref{eq:minDiv4}, which is a MC approximation of the cross-entropy minimization problem~\eqref{eq:LoptimProposDist_j}. Then, updated samples are drawn according to the new proposal $\nu_{j+1}$  (in step 3.2)  and related scores are computed (in step 3.3). The next quantile parameter $\rho_{j+1}$ and sample size $m_{j+1}$ are  tuned (in step 3.4) in order to satisfy an increase of at least $\delta>0$ of the quantile.  
After $J-1$ iterations, the quantile $\gamma( \hat \nu_{J-1},\rho_{J-1},\phi)$ is above the desired score $\gamma^{\star}$. The optimal proposal $\hat\nu_{A}^{\theta^\star}=\nu_J$ can then be computed (in step 4) using the original set~$A$.  Samples are drawn (in step 5) according to $\hat\nu_{A}^{\theta^\star}$ and related scores are computed (in step 6). The rare event probability is finally estimated by IS using these samples and scores (in step 7). 

The  free parameters of the algorithm are the initial sample size $m=m_0$, the initial quantile parameter $\rho=\rho_0 \in (0,1)$,  the minimal quantile increase value $\delta$ and the initial proposal $\nu_0$, which  is in most cases set to the distribution of $X$ (\ie $\nu_0=\mu$).

Let us make some important comments. The algorithm  imposes through condition \eqref{eq:condCEstandard}
  that quantiles will strictly increase at each of the algorithm iterations, implying that 
the estimated sequences of  $\nu_j$'s, $\rho_j$'s and $m_j$'s will be associated to a finite set of  stochastic events satisfying, as  the sets in~\eqref{eq:nestingOrig}, a nesting property
$
\hat A_1 \supseteq \hat A_2 \supseteq  \cdots \supseteq A.
$
This condition is used to prove the convergence of the algorithm as shown in Section~\ref{sec:theorie}.  

To guarantee  the validity of  \eqref{eq:condCEstandard} at each iteration, we use an upgraded version of the CE algorithm first proposed in \cite[Algorithm 5.2]{Homem-de-Mello02rareevent}. The difference with the basic CE method  \cite[Algorithm 2.1]{de2005tutorial} is that the algorithm adapts at the $j$-th iteration the parameter  $\rho_j$ and the sample size $m_j$ defining the  sequence of nested events~\eqref{eq:setAj}. This adaptation is presented in Algorithm~\ref{algo:1part1}. It consists in  checking if  $\rho_j$ can be tuned (in step 1) to meet the condition \eqref{eq:condCEstandard}. Otherwise  the sample size $m_j$ is increased (in step 3) by factor say $\beta=1.25$, \ie new samples are drawn (in step 2) and new scores are computed (in step 4)   until the condition is satisfied. The need for this adaptation step for convergence is justified by the asymptotic analysis exposed in Section~\ref{sec:theorie}.

Concerning complexity, the evaluation of  the  $\phi(z_i)$'s in Algorithm~\ref{algo:1}  or in  Algorithm~\ref{algo:1part1} requires $\mathcal{O}(mq^3)$ operations. It represents in general the most computational demanding step of the CE algorithm. In the favorable  situation where the  problem \eqref{eq:minDiv4} boils down to solving a linear system (\eg for $\mathcal{V}$ being the family of $p$-dimensional Gaussians, see Section~\ref{sec:paramGauss}), an update of the proposal density  requires  $\mathcal{O}(p^3)$ operations, yielding for $p \ll q$ the overall complexity $\mathcal{O}(Jmq^3)$. 
\subsection{Pre-conditioned CE Algorithm}
In the multi-fidelity method proposed in~\cite{peherstorfer2017multifidelity} and  detailed in Algorithm~\ref{algo:3}, the biasing densities are refined solving a sequence of ``low-fidelity CE methods''. For $k=1,\ldots, K+1$, the distribution $\nu^k$ is obtained (in step 1.2) solving the optimization problem related to a CE method defined (in step 1.1) with score function approximation $\phi^{(k)}\in \mathcal{M}$ instead of $\phi$, with $\mathcal{M}$ gathering in principle any type of surrogates. The strategy proposed by the authors is to use the distribution obtained as the solution of  $k$-th optimization problem of this sequence  as an initialization for the problem at level $k+1$. After running sequentially the $K$  ``low-fidelity  CE methods", the optimal proposal $\hat\nu_{A}^{\theta^\star}$ is obtained by running the standard CE method initialized with the   $K$-th optimization problem's solution $\nu^K$. The rare event probability is finally estimated by IS (in step 2).

We remark that at each of the $K$ iterations of Algorithm~\ref{algo:3}, the pre-conditioned method invokes Algorithm~\ref{algo:1}, which resorts itself to the adaptation step performed by Algorithm~\ref{algo:1part1}.  We notice that if one of the $K$  score function approximations is upper bounded by a value lower than  $\gamma^\star$,   a straightforward implementation of Algorithm~\ref{algo:3} will yield a non convergent estimator. Indeed,  in such a situation, there will exist an index $j'$ such that for any $j\ge j'$ the  condition $\gamma(\hat \nu_j, \rho_j , \phi^{(k)}) \ge \min\{\gamma^\star, \gamma(\hat \nu_{j-1},\rho_{j-1}, \phi^{(k)} )+\delta\}$ will never hold even for an infinite sample size. To avoid this pathological case, the adaptation of the sample size   is avoided in the $K$ first iterations of  Algorithm~\ref{algo:3} and replaced by an increment of $k$, \ie a  refinement from $\phi^{(k)}$ to $\phi^{(k+1)}$.

The main advantage  of the  pre-conditioned  CE algorithm is that, using this sound initialization,  Algorithm~\ref{algo:1} (where $\phi^{(k)}$ substitutes $\phi$) typically converges  at level $k$ in only a few iterations. As shown in~\cite{peherstorfer2017multifidelity} and mentioned in the following, some guarantees can be obtained under mild conditions on the number of iterations saved at each of the $K+1$ levels of Algorithm~\ref{algo:3}. Since this saving occurs in particular at the last level (\ie at $K+1$), it partially alleviates the computational bottleneck induced by calling too many times the high-fidelity model. The algorithm complexity remains of $\mathcal{O}(Jmq^3)$, as for the standard CE method.    However,  to obtain the initialization of Algorithm~\ref{algo:3} at the last level, $K$ optimization problems need first to be solved, each one of them  targeting a solution of the form~\eqref{eq:minDiv2} where $\phi^{(k)}$ substitutes $\phi$.    
In fact,  the proposed pre-conditioning does not necessarily guarantee that the global computational cost is lower or equal  to that of  the standard CE method.

\section{Recipe for Selecting Reduced Models in the CE Method}\label{sec:ourRecipes}
We propose in the following to incorporate the third ingredient and select the score  approximation in a hierarchy of pre-computed surrogates, at each level of the CE optimization process. 
\subsection{The Proposed Algorithm}
We present the proposed CE  method in Algorithm~\ref{algo:2}.
 It relies on  a {\it relaxed}  set of  events defined for $j\ge 0$ as 
  \begin{align}
  \hat A^{(k_j)}_{j+1}&=\{x\in \Rr^p : \phi^{(k_j)}(x) \ge \min\left(  \gamma( \hat \nu_{j},\rho_{j},\phi^{(k_j)}) -2 \alpha_{k_j}(\hat \nu_j), \gamma^\star+ \alpha_{k_j}(\hat \nu_j)\right) \},\label{eq:setAjs}
\end{align}
with the error bounds     \vspace{-0.5cm}
\begin{align}\label{eq:alphaHat}
 \alpha_k(\hat \nu_j)= c\max_{i=1,\cdots,m_j}\epsilon_{k_j}(z_i),
\end{align}
 where $c$ is the constant  in~\eqref{alpha_k} and  $z_1,\cdots,z_{m_{j}}$ are  i.i.d.  samples from $\nu_j$\footnote{We will see in Appendix~\ref{app:prelim1} that $\alpha_k(\hat \nu_j)$ converges as $m_j\to \infty$ to  $\alpha_{k_{j}}(\supp(\nu_{j}))$, \ie to a global error estimator for a distribution with infinite support. However, in practice, as $m_j$ is finite, $\alpha_k(\hat \nu_j)$  is rather a local error bound computed with samples drawn in the region where distribution  $\nu_j$ has a lot of mass.}. 
As described below, the algorithm determines the sequences of  $k_j$'s,  $m_{j}$'s, $\rho_{j}$'s and $ \nu_{j}$'s.
 The idea behind Algorithm~\ref{algo:2} is to compute, as for the standard CE method, a sequence of $(\nu_j, \rho_j, m_j)$'s satisfying~\eqref{eq:condCEstandard} or equivalently a sequence of nested events  $ \hat  A_{j+1} $  including the target set of rare events $A$. This computation should rely as much as possible  on the score function approximations and be controlled by their error estimates.  In other words, the philosophy is to exploit the approximations and error bounds to  adapt sequences of  $k_j$'s,  $m_j$'s, $\rho_{j}$'s and $ \nu_{j}$'s so that $\hat A^{(k_{j})}_{j+1}$ is the smallest set  built with the score approximation $ \phi^{(k_j)}$ which contains  $ \hat  A_{j+1}$: for any $z \in \Rr$, the set   $\{x\in \Rr^p :   \phi^{(k_j)}(x) \ge z, x \in \hat  A_{j+1}\}$ should contain or be equal to $\hat A^{(k_{j})}_{j+1}$.

\subsubsection{General structure}

Algorithm~\ref{algo:2} is pretty much similar to  Algorithm~\ref{algo:1}. Its 7-steps structure differs slightly from the standard CE method, by the use of selected score approximations.  
We hereafter comment each step of the algorithm.  At the beginning, $m$ samples are drawn (in step 1) from the initial proposal $\nu_0$  and related score approximations and error bounds are computed (in step~2). Then, the algorithm estimates (in step 3) a sequence of proposals $\{\nu_1,\ldots, \nu_{J-1}\}$, and  related sequences of  quantile parameters $\{\rho_1,\ldots, \rho_{J-1}\}$, sample sizes $\{m_1,\ldots, m_{J-1}\}$ and approximation levels $\{k_1,\ldots, k_{J-1}\}$,  until the quantile computed with approximation $\phi^{(k_j)}$ reaches the desired score $\gamma^{\star}$ augmented by the error bound $\alpha_{k_j}(\hat \nu_j)$ (as proved in Section~\ref{sec:theorie}, the addition of this term guarantees that the  quantile  computed with  the original $\phi$ is above $\gamma^{\star}$). 
Let us detail the building of these sequences.   At the $j$-th iteration, given $\rho_j \in (0,1)$, a sample size $m_j$, a proposal $\nu_j$ and an approximation level $k_j$, the algorithm evaluates the empirical $(1-\rho_j)$-quantile  $ \gamma(\hat \nu_j, \rho_j, \phi^{(k_j)} )$ using  samples  $z_1,\cdots,z_{m_j}$ from $\nu_j$. This quantile together with the related error bound define  the stochastic event 
$\hat A^{(k_{j})}_{j+1}$  through \eqref{eq:setAjs}.
 Given  the set $\hat A^{(k_{j})}_{j+1}$ and the proposal  $\nu_{j}$,  a refined distribution $\nu_{j+1}$ is estimated (in step 3.1) by solving ~\eqref{eq:minDiv4_}, which is a MC approximation of the cross-entropy minimization problem~\eqref{eq:LoptimProposDist_j} using the  $\phi^{(k_j)}$ in place of the original score function $\phi$. Then, updated samples are drawn according to the new proposal $\nu_{j+1}$  (in step 3.2)  and related score approximations and error bounds are computed (in step 3.3). The next quantile parameter $\rho_{j+1}$ and sample size $m_{j+1}$  are  tuned (in step 3.4), together with the approximation level $k_{j+1}$, in order to satisfy a quantile increase of at least $\delta+2\alpha_{k_j}(\hat \nu_j)$ (as shown in Section~\ref{sec:theorie}, the addition of the term $2\alpha_{k_j}(\hat \nu_j)$  guarantees that the  quantile  computed with  the original $\phi$ increases of $\delta$).
After $J-1$ iterations, the quantile $\gamma( \hat \nu_{J-1},\rho_{J-1},\phi^{(k_{J-1})})$ is above the  score $\gamma^{\star}+\alpha_{k_{J-1}}(\hat \nu_{J-1})$.  The end of  the algorithm is identical to the standard CE method: the optimal proposal $\hat\nu_{A}^{\theta^\star}=\nu_J$ is computed (in step 4) using the original set~$A$;  samples are drawn (in step 5) according to $\hat\nu_{A}^{\theta^\star}$ and related scores are computed (in step 6); the rare event probability is finally estimated by IS using these samples and scores (in step 7). 

The  free parameters of the algorithm are the same as for the standard CE method, supplemented by the hierarchy $\mathcal{M}$ of score approximations.

\begin{algorithm}[H]
\begin{algorithmic}[0]
\State \textbf{input parameters}: ($m$, $\rho$, $\delta$, $\nu_0$, $\mathcal{M}$)
 \State {{\it \bf 1}- Draw  $m$ i.i.d.  samples $z_1,\cdots,z_{m}$ from $\nu_0$.}
\State {{\it \bf 2}- Compute $\phi^{(1)}(z_i)$'s and $ \alpha_{1}(\hat \nu_{0})$.}
\State {{\it \bf 3}- Set  $j=0$, $m_0=m$,  $\rho_0=\rho$ and $k_0=1$}.
 \While{
  $\gamma( \hat \nu_{j},\rho_{j},\phi^{(k_j)}) < \gamma^{\star} + \alpha_{k_j}(\hat \nu_j)$
 }
\State{{\it \bf 3.1}-  Set $ j=j+1.$ and solve\vspace{-0.5cm}
 \begin{align}\label{eq:minDiv4_}
&\nu_j \in  \argmax_{\nu^\theta \in \mathcal{V}} \frac{1}{m_{j-1}} \sum_{i=1}^{m_{j-1}} \unit_{\hat A^{(k_{j-1})}_j}(z_i) \frac{\mu(z_i)}{ \nu_{j-1}(z_i)} \ln \nu^\theta(z_i).
\end{align}}
\State {{\it \bf 3.2}- Draw $m_{j-1}$  i.i.d.  samples $z_1,\cdots,z_{m_{j-1}}$ from $\nu_j$.}
\State {{\it \bf 3.3}- Compute  $\phi^{(k_j)}(z_i)$'s and $ \alpha_{k_j}(\hat \nu_{j})$'s.}
\State {{\it \bf 3.4}- Using Algorithm~\ref{algo:2part1}, adapt  $m_j$, $\rho_j$, and $k_j \in \{k: \phi^{(k)}\in \mathcal{M}\}$ \\\hspace{1.2cm}starting from
  $m_j=m_{j-1}$, $\rho_j=\rho_{j-1}$, $k_j=k_{j-1}$ such that
 \begin{align}
 &\gamma(\hat \nu_j, \rho_j ,\phi^{(k_j)}) \ge \bar \gamma=\min\{\gamma^\star+ \alpha_{k_j}(\hat \nu_j) , \tilde \gamma\},\label{eq:deltaGamma}\\
 &  \alpha_{k_{j}}(\hat \nu_{j}) \le   \alpha_{k_{j-1}}(\hat \nu_{j-1}),\label{eq:deltaError}
\end{align}
 \hspace{1.2cm} where   $\tilde \gamma$=$  \gamma(\hat \nu_{j-1},\rho_{j-1},\phi^{(k_{j-1})} )  +2   \alpha_{k_{j-1}}(\hat \nu_{j-1}) + \delta.$
 }
 \EndWhile
  \State{{\it \bf 4}- Set  $ J=j+1$ and solve~\eqref{eq:minDiv4_} with $\hat A^{(k_{j-1})}_j=A$ }.

\State{{\it \bf 5}- Draw $m_{J-1}$  i.i.d.  samples $z_1,\cdots,z_{m_{J-1}}$ from $\nu_J$.}
\State{{\it \bf 6}- Compute  $\phi(z_i)$'s.}
\State {{\it \bf 7}-  Set $\hat\nu_{A}^{\theta^\star}=\nu_J$ and compute $  p^{IS}_{A,\hat\nu_{A}^{\theta^\star}}$  using~\eqref{eq:IS}  for $\nu=\hat\nu_{A}^{\theta^\star}$}.
\State \textbf{outputs}:  ($\hat\nu_{A}^{\theta^\star}$, $  p^{IS}_{A,\hat\nu_{A}^{\theta^\star}}$).

\end{algorithmic}

\caption{ - CE algorithm with score approximation  selection \label{algo:2}}
\end{algorithm}
\begin{algorithm}[H]
\begin{algorithmic}[0]
\State \textbf{inputs}: ($m_{j}$, $\rho_{j}$, $k_j$,  $z_i$'s,  $\beta$)
\While{\eqref{eq:deltaError} and~\eqref{eq:deltaGamma} do not hold}
	\If{ $k_j=K+1$}
		\State{{\it \bf 1}- Adapt $m_j$ and $\rho_j$ using Algorithm~\ref{algo:1part1}}.
	\Else
		\State{{\it \bf 2}-  Compute $\varpi={\underline{\rho}}(\hat \nu_j )- \bar \eta_{{\bar \gamma},k_j}(\hat \nu_j )$ using definition~\eqref{eq:part1} }.
		\If{$\varpi>0$ } 
			\State {{\it \bf 2.1}- 
			 Adapt $m_j$ and $\rho_j$ using Algorithm~\ref{algo:1part1} with $\phi^{(k_j)}$ in place of $\phi$.}
		\Else 
			\State {{\it \bf 2.2}- Select $k_j=k_j+1$ and compute  $\phi^{(k_j)}(z_i)$'s and $ \alpha_{k_j}(\hat \nu_{j})$'s.}
		\EndIf
	\EndIf
\EndWhile
\State \textbf{ouputs}: ($m_{j}$, $\rho_{j}$, $k_j$, $z_i$'s)
\end{algorithmic}
\caption{ -  Adaptation of  $m_j$, $\rho_j$ and selection of $k_j$ \label{algo:2part1}}
\end{algorithm}
 Finally, let us point out that problem~\eqref{eq:minDiv4_} may be ill-conditioned, \ie that the solution $\nu_j$ might not be unique. The non-uniqueness might be in certain circumstances only the consequence of an insufficient sample size, the circumstances being determined by the parametric distribution family $\mathcal{V}$. This issue is discussed in the case of the $p$-dimensional Gaussian family in Section~\ref{sec:numEvalAlgo}. 

 \subsubsection{Selection of Score  Approximations} \label{sec:selection}
 The adaptation step is driven by Algorithm~\ref{algo:2part1}, in place of Algorithm~\ref{algo:1part1}   in the standard CE method, and  includes  the selection of the score approximation. In the case $k_j=K+1$, \ie no score approximation are used, Algorithm~\ref{algo:2part1} boils down to Algorithm~\ref{algo:1part1} (step 1). In other situations, the algorithm checks if  the current approximation $\phi^{(k_j)}$ allows for the existence of a non-zero $\rho_j$  satisfying the nesting property and agreeing with an error tolerance, taking the form of conditions \eqref{eq:deltaGamma} and~\eqref{eq:deltaError}. As shown  in Section~\ref{sec:TheoremProofPart1}  these two conditions guarantee asymptotically an increase of the quantile computed with the original high-fidelity score of at least~$\delta$.  
 
 However, it is important to remark that at  the current approximation level $k_j$, the condition  $\gamma(\hat \nu_j, \rho_j, \phi^{(k_j)} )\ge \bar \gamma$  may in some cases only be satisfied for $\rho_j=0$, even if we increase the sample size $m_j$. This situation occurs in particular in the case where $\phi^{(k_j)}$ under-estimates systematically  the score function.  However, we do not know {\it a priori} if we are in such an undesirable  situation.  
Rather than increasing
 $m_j$ and trying to obtain in vain an inaccessible non-zero value    $\rho_j$, it is more cautious to use the current sample size  and select the next possible approximation level in $\mathcal{M}$, for which there exists $\rho>0$ such that the condition  $\gamma(\hat \nu_j, \rho, \phi^{(k_j)} )\ge \bar \gamma$ stands. 
 
  Ideally, in order to make the selection of $k_j$ more robust, we should check the  existence of $\rho>0$ such that  $\gamma(\nu_j, \rho, \phi^{(k_j)} )\ge \bar \gamma$ holds, 
 and not only rely on the samples of the empirical distribution $\hat \nu_j$. We derive hereafter a  worst-case sufficient condition for the existence of such an ideal condition, worst-case meaning that the approximated score under-estimates systematically the high-fidelity score of a value equal to the error bound.  The worst-case sufficient condition is derived from  the assumption that $\rho_{\bar \gamma}=\langle   \unit_{\phi(\cdot) \ge \bar \gamma},  \nu_j \rangle>0$ (which is a mild assumption used in Section~\ref{sec:asssumptions} to prove convergence).  Indeed as shown in Appendix~\ref{app:proofProp1},  a  sufficient condition  in this worst-case scenario  is that $\rho_{\bar \gamma}> \eta_{\bar \gamma,k_j}$, where $\eta_{\bar \gamma,k_j}$ is the probability that the approximated score falls into the interval $[ \gamma(  \nu_j,\rho_{\bar \gamma},\phi^{(k_j)}), \gamma(  \nu_j,\rho_{\bar \gamma},\phi^{(k_j)})+ \alpha_{k_j}( \nu_j)]$.

 In practice,  $\rho_{\bar \gamma}$ and  $\eta_{\bar \gamma,k_j}$ are both inaccessible as they  depend on   the high-fidelity  score $\phi$ on one side, and on the other side  on intractable integrals. 
 To overcome this issue, we first derive a lower bound  on $\rho_{\bar \gamma}-\eta_{\bar \gamma,k_j}$  using the following remark, and  design a tractable  plug-in estimator of this bound. \\
 \begin{remark}\label{rem:1}
 Let  $\alpha_k( \nu)$ and $\alpha_\ell( \nu) $ be error bounds such that $\alpha_k( \nu) \le \alpha_\ell( \nu) $. From 
$
 |\phi^{(k)}(x)-\phi(x)| \le\alpha_\ell (\nu),$ we deduce  $ \langle\unit_{\phi(\cdot)  < s-\alpha_\ell( \nu)} , \nu \rangle \le \langle\unit_{\phi^{(k)}(\cdot)  < s} , \nu \rangle \le \langle\unit_{\phi(\cdot)  < s+\alpha_\ell( \nu)} , \nu \rangle,$ 
 which in turns leads to  
$  | \gamma( \nu,\rho,\phi^{(k)})-\gamma( \nu,\rho,\phi) | \le \alpha_\ell( \nu)$  for any $\rho \in (0,1)$.
\end{remark}\\

Therefore,  a lower bound for $\rho_{{\bar \gamma}}$ and an upper bound for  $\eta_{{\bar \gamma},k_j}$ are available, from which we design  the
related plug-in estimators, respectively  defined as
  \begin{align}\label{eq:part1}
  {\underline{\rho}}_{{\bar \gamma}}(\hat \nu_j )&=\langle   \unit_{\phi^{(k_j)}(\cdot) \ge {\bar \gamma}+ \alpha_{k_{j}}(\hat \nu_{j})}, \hat \nu_j \rangle,\nonumber \\
\bar \eta_{{\bar \gamma},k_j}(\hat \nu_j )&=\max_{\gamma'\in [ \gamma_b , \gamma_u ]}\langle  \unit_{{\phi^{(k_j)}(\cdot) \in [\gamma',\gamma'+\alpha_{k_{j}}(\hat \nu_{j})}]}, \hat \nu_j\rangle,
    \end{align}
with $ \gamma_b = \gamma(   \hat \nu_j ,{\bar{\rho}_{{\bar \gamma}}(\hat \nu_j )}, \phi^{(k_j)})$,  $ \gamma_u = \gamma( \hat  \nu_j ,{\underline{\rho}}_{{\bar \gamma}}(\hat \nu_j ), \phi^{(k_j)})$ and 
$ {\bar{\rho}_{{\bar \gamma}}(\hat \nu_j )}=\langle   \unit_{\phi^{(k_j)}(\cdot) \ge {\bar \gamma}- \alpha_{k_{j}}(\hat \nu_{j})}, \hat \nu_j \rangle$.
We observe that the positivity of  $\varpi={\underline{\rho}}_{{\bar \gamma}}(\hat \nu_j )- \bar \eta_{{\bar \gamma},k_j}(\hat \nu_j )$ implies in particular the validity of condition \eqref{eq:deltaGamma}, making the first condition more demanding than the second. The criterion for the selection of the approximation level  is thus more exigent than the quantile increase condition used once an approximation level has been selected.

Therefore, computing $\varpi$  (in step 2) and checking its positivity  constitutes our criterion for the selection of the next approximation level $k_j$. 
More precisely, if $\varpi>0$ and the level $k_j$  is selected, the algorithm adapts (in step 2.1)  $\rho_j$ and $m_j$ such that $\gamma(\hat \nu_j, \rho_j, \phi^{(k_j)} )\ge \bar \gamma$  using Algorithm~\ref{algo:1part1} with $\phi^{(k_j)}$ in place of $\phi$. In the case where $\varpi\le 0$,  $k_j$ is increased (in step 2.2) to the next possible value in $\mathcal{K}$, \ie a refined score function approximation~$\phi^{(k_j)}$ is selected in $\mathcal{M}$. This refinement process continues  until  the  condition $\varpi> 0$ is met  or the highest level $k_j=K+1$  (for which $\phi^{(K+1)}=\phi$) is reached.

\subsection{Computational Saving}

In comparison to state-of-the-art, the main innovation of Algorithm~\ref{algo:2} is that it selects at each of its iteration the score function approximation in a hierarchy of pre-computed surrogates to meet the need for current accuracy. More precisely, the proposed algorithm uses score function approximations 
  directly within the core of the CE method, and furthermore  parsimoniously guided by  error estimates (in particular  all score  approximations are not systematically used).

As for the standard CE algorithm, the computation   of the high-fidelity  score function or of its approximations  represents  the most computational intensive steps. Of course, the evaluation of the high-fidelity score function for a  set of samples is  irrevocable  in the final step  for the estimation of the rare event  probability. 
  Nevertheless, in the previous iterations of the proposed algorithm,  surrogates are chosen sparingly, in order to significantly reduce the computational burden.  As shown  in the analysis of Section~\ref{sec:theorie}, the computational gain is not at the expense  of a deterioration of  the  sequence of proposal densities. Indeed, the sequence of  $\nu_j$'s generated by the proposed  algorithm are asymptotically guaranteed  to satisfy the quantile increase  condition~\eqref{eq:condCEstandard} and converge to the optimal density~\eqref{eq:minDiv2}, just as for the standard CE algorithm.  

The algorithm complexity remains of $\mathcal{O}(Jmq^3)$, as for state-of-the-art  methods.
In theory, we note that if the bound on the maximum number of iterations of  Algorithm~\ref{algo:2} and Algorithm~\ref{algo:1} are identical (and we will see that this is the case in the next section), then the worst computational load involved by the former is guaranteed to be lower or equal to the one involved by the latter.  
In practice, we expect a large number of iterations to be performed with the reduced models, \ie requiring  only $\mathcal{O}(md_k^3)$ operations, in the usual setting where $r \le d_k$ with  $r$ the number of affine parameters. 

Let us point out that the error bounds are computed in $\mathcal{O}(mr^2d_k^2)$ operations. Therefore, this does not increase significantly the computational cost as long as $r^2 \ll d_k$. This regime is in  general the only one suitable for importance sampling, which becomes very challenging in high dimensions as pointed out in~\cite{au2003important}. Besides, we remark that the evaluation of $\bar \eta_{{\bar \gamma},k_j}(\hat \nu_j )$ at the core of the the selection of the approximation level involves a simple $1$-dimensional minimization. Once the score function approximations have been computed, it can be solved with a negligible  computational load using standard optimization techniques.

Finally, we remark that making an on-line adaptation of the directions (in addition of the dimension) of the RB spaces is prohibitive if we want to maintain an on-line cubic complexity in $d_k$, independent of the ambient dimension $q$. Indeed,  updating a direction in the reduced model  would require  $\mathcal{O}(r d_k q^2)$ operations,   see Section~\ref{sec:2ndIng}. Nevertheless,  the  algorithm's efficiency  could be increased reconsidering the off-line/on-line decomposition paradigm, designing so-called  {\it adaptive} reduced models built by  enrichment of the RB spaces as proposed in~\cite{peherstorfer2015online,Chen:196807}.
This possible  enhancement of the algorithm is however out of the scope of the present work.

\section{Asymptotic Analysis}\label{sec:theorie}
We provide  theoretical guarantees  on the convergence of the proposed algorithm. We begin by introducing assumptions and reviewing state-of-the-art results.   

\subsection{Assumptions}\label{sec:asssumptions}

We use the following  assumptions, as done in~\cite{Homem-de-Mello02rareevent}. \\

\noindent
\textbf{Assumption A:}
 For any $\nu^\theta \in \mathcal{V}$, $\textrm{supp}(\mu) \subseteq  \textrm{supp}(\nu^\theta)$.\\

In words, Assumption A ensures that the domination relation  $\textrm{supp}(\mu) \subseteq  \textrm{supp}(\nu_A^{\theta^\star})$, satisfied by the zero-variance density $\nu_A^{\theta^\star}$ given by~\eqref{eq:minDiv2} is also  satisfied for any feasible $\nu^\theta \in \mathcal{V}$. This assumption is trivially satisfied if densities in $\mathcal{V}$ have an infinite support.

We remark that  Assumption A implies that  $\langle \unit_{A}, \nu^\theta \rangle >0 $ for any $\nu^\theta \in \mathcal{V}$  as long as $\langle \unit_{A}, \mu \rangle >0 $, \ie  as long as the sought rare-event probability is non-zero. 
We mention that this assumption could be relaxed as done in~\cite{Homem-de-Mello02rareevent}. \\

\noindent
\textbf{Assumption B:}
We assume that
\begin{itemize}
\item[{\it i)}]  the set $\Theta$ is compact; 
\item[{\it ii)}] for almost every $\theta$, the function $ \ln \nu^\theta$ is continuous on $\Theta$; 
\item[{\it iii)}]  there exists a function $h:\Rr^p \to \Rr$ such that $\langle h, \mu\rangle <\infty$ and $| \ln \nu^\theta(x) | \le h(x)$ for all $x\in \Rr^p$ and $\nu^\theta \in \mathcal{V}$;\\
\end{itemize}

The  properties {\it i)}, {\it ii)} and {\it iii)} are satisfied by  numerous family of distributions. To illustrate that, consider the Gaussian family
$
\mathcal{V}=\{\nu^\theta: \nu^\theta(x)= {1}/{\theta}e^{(-{\|x\|^2_2}/{\theta})}, \theta \in \Theta \subset \Rr^+ \},
$
 where $\Theta$ is a compact excluding zero. In this case, we  verify that Assumption B holds, and in particular  there exists a quadratic  function $h$ satisfying~{\it iii)}. \\

%

\subsection{Previous Results}

We hereafter expose the convergence result for the standard CE method~\cite{Homem-de-Mello02rareevent}. \\

\begin{theorem}[Homem de~Mello and Rubinstein, 2002]\label{theorem:0}
Suppose that Assumptions A and B hold.  Then, in  Algorithm~\ref{algo:1} as $m_j\to \infty$,  
\begin{itemize}
 \item  the biasing distribution ${\hat\nu_{A}^{\theta^\star}}$  converges almost surely to~\eqref{eq:minDiv2} in a number of iterations lower or equal to\vspace{-0.15cm}
 \begin{align}\label{eq:Jmax1}
\hat J_{\max}= \lceil  \frac{\gamma^\star- \gamma( \hat \nu_{0},\rho_{0},\phi)}{\delta}\rceil +1;
\end{align}

\item  the  SCV of the estimator~\eqref{eq:IS} is
\begin{itemize}
 \item  at the $j$-th iteration using $\nu=\nu_j$ and assuming $\nu_{\hat A_j}^{\star} \in \mathcal{V}$ 
 $$\frac{1}{p_A^2}{\mathbb{E}[(  p^{IS}_{A,\nu_j}-p_A)^2]}= \frac{1}{m_jp_{A}}{p_{\hat A_j\setminus A}};$$  

\item equal to zero at the $J$-th iteration using $\nu=\nu_J$ and assuming $\nu_{A}^{\star} \in \mathcal{V}$.\\
 \end{itemize}
 \end{itemize}
%
%

\end{theorem}

The almost sure convergence in at most $\hat J_{\max}$ iterations  is proven in \cite{Homem-de-Mello02rareevent}.  The result on the estimator SCV after the $j$-th iteration  is straightforward, as detailed in Appendix~\ref{app:0} and obvious after the $J$-th iteration  as the algorithm sets $\hat A_j=A$ implying that $p_{\hat A_j\setminus A}=0$. We see that, the smaller the set ${\hat A_j} \setminus {A}$, the smaller the  variance. Therefore, it is tempting to build a sequence  $\hat A_j$ tending quickly to $A$ which would yield a drop-off on the estimator variance. However, the drawback of such a construction is that this approach can be   expensive since it generally requires  large values for the $m_j$'s. Indeed,  the value of $m_j$ may need in this situation to be increased significantly until it fulfils   condition~\eqref{eq:condCEstandard}. In fact, there is a tradeoff between the value of $m_j$'s and  parameter $\delta$ which imposes a minimal speed of convergence of the sequence of $\hat A_j$'s towards $A$.

We mention that the authors in \cite[Proposition 1]{peherstorfer2017multifidelity} slightly enhance the first statement of Theorem~\ref{theorem:0}. They prove that, under mild conditions, the bound on the maximum number of iterations becomes lower or equal to $\hat J_{\max}$ when their  multi-fidelity approach is used to initialize the CE method. 
 The assumption guaranteeing that the  bound on the maximum number of iterations is lowered at  level $k$ (and in particular at the last level $K+1$ using the high-fidelity model) is that 
$
\langle \unit_{\phi^{(k)}(\cdot) \ge \gamma(\nu^{k-1}, \rho ,\phi^{(k)}) },\nu^{k-1} \rangle \ge \langle \unit_{\phi^{(k)}(\cdot) \ge \gamma(\nu^{k-1}, \rho ,\phi^{(k)}) },\mu\rangle.
$
In practice, this assumption holds if we can  verify  conditions on the score function approximation regularity and on its cumulative distribution with respect to the biasing densities in $\mathcal{V}$, see \cite[Assumption 1 - 2]{peherstorfer2017multifidelity}. 
Nevertheless,
there exists  no theoretical guarantees that the $K+1$ level of the pre-conditioned method are more  efficient than the standard  CE method. 

\subsection{Our  Result}

We now present the theoretical guarantees obtained with the proposed method in the asymptotic regime.  \\

\begin{theorem}\label{theorem:1}
Suppose that Assumptions A and B hold.  Then,  in Algorithm~\ref{algo:2} as $m_j\to \infty$,  
\begin{itemize}
 \item   the biasing distribution ${\hat\nu_{A}^{\theta^\star}}$ converges almost surely to~\eqref{eq:minDiv2} in at most $\hat J_{\max}$ iterations, as for the standard  CE method; 
\item   the  SCV of the estimator~\eqref{eq:IS} is
\begin{itemize}
 \item at the $j$-th iteration  using $\nu=\nu_j$  and assuming $\nu_{\hat A_j^{(k_{j-1})}}^{\star} \in \mathcal{V}$
$$
\frac{1}{p_A^{2}}{\mathbb{E}[(  p^{IS}_{A,\nu_j}-p_A)^2]}= 
\frac{1}{m_jp_A}\left(p_{\hat A_j\setminus A}+p_{\hat A^{(k_{j-1})}_j \setminus \hat A_j} \right),
$$ 
which is the minimal SCV achievable among all  feasible   $\hat A^{(k_{j-1})}_j$'s;
\item   equal to zero   at the $J$-th iteration using $\nu=\nu_J$ and assuming  $\nu_{A}^{\star} \in \mathcal{V}$, as for the standard  CE method.\\
  \end{itemize}
 \end{itemize}
\end{theorem}

In the theorem, we call  the tuple $(\nu_j, \rho_j, k_j )$,  or equivalently  the set $\hat A^{(k_j)}_{j+1}$ given in~\eqref{eq:setAjs}  {\it feasible}  if and only if  $(\nu_j, \rho_j)$ satisfies~\eqref{eq:condCEstandard}.
The first statement of this theorem  proves the almost sure convergence of our algorithm towards the optimal biasing distribution in at most $\hat J_{\max}$ iterations, as  the standard CE algorithm. In this worst case scenario of $\hat J_{\max}$ iterations, we deduce that Algorithm~\ref{algo:2} will use the same or lower computation power than  the standard CE method, since at each iteration  the score function  can be exchanged for an approximation of  lower complexity.

The second statement of this theorem  shows that  the  squared coefficient of variations of the estimator~\eqref{eq:IS} obtained with  the biasing distribution  at the $j$-th iteration of Algorithm~\ref{algo:1} and Algorithm~\ref{algo:2} are identical up to an additional term proportional to the probability of the set ${\hat A^{(k_{j-1})}_j \setminus \hat A_j} $.  Furthermore, the statement claims that this probability is the minimal achievable among all feasible $\hat A^{(k_{j-1})}_j$. In other words, given the score functions approximation $\phi^{(k_{j-1})}$ and the worst-case error 
$ \alpha_k(\hat \nu_j)$, the choice of the set  $\hat A^{(k_{j-1})}_j$ is optimal  at the $j$-th iteration in the sense that no other feasible set can  yield  a lower variance of the IS estimator. 

According to this second statement,  the use of a relaxed set  containing  $\hat A_j$  (even if chosen optimally in the feasible set)  instead of $\hat A_j$ itself (as done in the standard CE algorithm) implies unavoidably an additional term in the SCV. 
 However, it is important to point out that the choice of the relaxed set at the $j$-th iteration of the algorithm should not necessarily be guided by  obtaining   the lowest  achievable SCV (as done in the  standard CE algorithm). The criterion to be minimized is rather the global computational cost for the estimation of the target: the optimal  biasing density~$\nu_J$. 
  Indeed, whatever the chosen path for the $\nu_j$'s  in the distribution space from $\nu_0=\mu$ to $\nu_J$ and whatever their associated SCV's,  $\nu_J$ will always be associated with the same SCV,  independently of this path and the involved algorithm. Therefore, the objective is to obtain the optimal density with a minimum of effort. In this perspective,  the proposed algorithm 
builds  at the $j$-th iteration  the relaxed set $\hat A^{(k_{j-1})}_j \supseteq \hat A_j$ for inference of  $\nu_j$  so that 1) the standard quantile increase condition~\eqref{eq:condCEstandard} holds, 2) the estimation involves the score function approximation of lowest complexity.
Thanks to these two properties,  the optimal density~\eqref{eq:minDiv2} is reached by the proposed algorithm in at most $\hat J_{\max} $ iterations, just as for the standard CE algorithm,  but with possibly a minimal computation burden.


\subsection{Proof of Theorem~\ref{theorem:1}}\label{sec:TheoremProof}

\subsubsection{First Statement: Convergence}\label{sec:TheoremProofPart1}

 \proof{This statement is a straightforward consequence of the two  propositions given in this section. $\square$}\\
 

\begin{proposition}\label{cor:22}
Under assumptions  A and B,  Algorithm~\ref{algo:2part1} selects   $\{\rho_j\}_{j=1}^{J}$, $\{k_j\}_{j=1}^{J}$ and $\{m_j\}_{j=1}^{J}$ such that the sequence $\{(\nu_j,\rho_j)\}_{j=1}^{J}$ produced by Algorithm~\ref{algo:2} satisfies  almost surely  as $m_j \to \infty $ the condition \begin{align}\label{eq:nestingNotRelax}
\gamma( \nu_j, \rho_j ,\phi) \ge \min\{  \gamma^\star,  \gamma(\nu_{j-1}, \rho_{j-1} ,\phi)  +\delta \}.
\end{align}  \vspace{-0.4cm}
\end{proposition}

\proof{We begin by verifying that Algorithm~\ref{algo:2part1}  searches at each iteration the  smallest    $k_j$ such that  \eqref{eq:deltaError} holds and   $\varpi>0$.  
Note that the latter condition holds only if  ${\underline{\rho}}(\hat \nu_j )>0$, which in turns implies   the existence of $\rho_j>0$ such that \eqref{eq:deltaGamma}.
 The proof of Proposition~\ref{cor:22}   then relies  on the two following arguments. 
 On the one hand, statements {\it i)} and {\it ii)}  in Lemma~\ref{prop:cvMC} show 
 that, for $m_j$ large enough,  conditions  \eqref{eq:deltaGamma} and \eqref{eq:deltaError} imply  their deterministic counterparts\vspace{-0.15cm}
\begin{align}
&\gamma( \nu_j, \rho_j ,\phi^{(k_j)}) \ge \min\{  \gamma^\star+\alpha_{k_j}(\nu_j), \tilde \gamma_{\ell} \},\label{eq:nestingRelax}\\
&\alpha_{k_j}(\nu_j) \le \alpha_\ell(\nu_{j-1}),\label{eq:nestingRelaxCond2}\vspace{-0.15cm}
\end{align}
where $\tilde \gamma_{\ell}= \gamma(\nu_{j-1}, \rho_{j-1} ,\phi^{(\ell)}) +2\alpha_\ell(\nu_{j-1}) +\delta$.
On the other hand, using   Remark~\ref{rem:1} to  major $\gamma( \nu_{j},\rho_{j},\phi^{(k_j)})$ and minor $ \gamma( \nu_{j-1},\rho_{j-1},\phi^{(\ell)})$, it is straightforward to verify that  conditions \eqref{eq:nestingRelax} and \eqref{eq:nestingRelaxCond2} imply  that $(\nu_j,\rho_j)$ satisfies~\eqref{eq:nestingNotRelax}.
 $\square$}
\\

In words, this first proposition shows  that the sequence $\{(\nu_j,\rho_j)\}_{j=0}^{J}$  built by the proposed algorithm satisfy asymptotically the deterministic counterpart of the quantile  increase condition \eqref{eq:condCEstandard} used in the standard CE method. 
In particular, it shows that  events
$
 A_1 \supset  A_2 \supset  \cdots  \supset A_J  = A,
$
 are  nested according to~\eqref{eq:nestingNotRelax} where $J< J_{\max} $.\\

\begin{proposition}\label{cor:33}
If  the sequence $\{(\nu_j,\rho_j)\}_{j=1}^{J}$ satisfies~\eqref{eq:nestingNotRelax} then under Assumption B,  the biasing distribution ${\hat\nu_{A}^{\theta^\star}}$ given by Algorithm~\ref{algo:2} converges to the optimal one~\eqref{eq:minDiv2}  almost surely as $m_j\to \infty$  in $J\le \hat J_{\max}$  iterations. \\
\end{proposition}

\proof{
To show this result we  introduce the deterministic counterpart of problem~\eqref{eq:minDiv4_}, that is\vspace{-0.2cm}
 \begin{align}\label{eq:minDiv4_continue}
   \argmax_{\nu^\theta \in \mathcal{V}}\langle  \unit_{ A^{(k_{j-1})}_j}  \ln \nu^\theta , \mu\rangle,
\end{align}
where $A^{(k_{j-1})}_{j}=$\vspace{-0.2cm}
 \begin{align*}
   \{x\in \Rr^p :
    \phi^{(k_{j-1})}(x) \ge \min \left(  \gamma( \nu_{j-1},\rho_{j-1},\phi^{(k_{j-1})}) -2\alpha_{k_{j-1}}(\nu_{j-1}), \gamma^\star+\alpha_{k_{j-1}}(\nu_{j-1}) \right) \}. 
  \end{align*}

We  show  in  Appendix~\ref{app:prelim1} that  under Assumption B, the distribution $\nu_j$ in Algorithm~\ref{algo:2} (the solution~\eqref{eq:minDiv4_})  converges  almost surely as $m_j \to\infty$  to the solution~\eqref{eq:minDiv4_continue}. This result corresponds to statement {\it iii)} of Lemma~\ref{prop:cvMC}. 
 Moreover, because the sequence $\{(\nu_j,\rho_j)\}_{j=1}^{J}$ satisfies~\eqref{eq:nestingNotRelax} according to Proposition~\ref{cor:22}, we have $\gamma( \nu_{j-1}, \rho_{j-1} ,\phi) \le \gamma( \nu_{j}, \rho_{j} ,\phi) +\delta$ for $j=1,\cdots, J$. And at the $(J-1)$-th iteration we have $\gamma( \nu_{J-1},\rho_{J-1}, \phi) \ge \gamma^\star$.
Thus, step 4 of Algorithm~\ref{algo:2} provides the optimal distribution $\argmax_{\nu^\theta \in \mathcal{V}}\langle  \unit_{ A}  \ln \nu^\theta , \mu\rangle$. Finally,  it is clear from~\eqref{eq:nestingNotRelax} and the almost sure convergence of quantile and error bound approximations  (statements {\it i)} and {\it ii)} of  Lemma~\ref{prop:cvMC})  that   for $m$ large enough, the number of iterations of the algorithm, \ie $J=\min\{j\ge 1:  \gamma( \hat \nu_j, \rho_j ,\phi^{(k_j)}) \ge  \gamma^\star+ \alpha_{k_j}(\hat \nu_j)\}+1,$
is upper bounded  by 
$ \hat J_{{\max}}$ defined in~\eqref{eq:Jmax1},  \ie  $J \le \hat J_{{\max}}$.
$ \square$ }\\

Relying on 
Remark~\ref{rem:1} in Section~\ref{sec:selection},  we can compute in practice an upper bound on this maximum number of iteration
\begin{align*}
 \hat J_{{\max}} \le \lceil  \frac{\gamma^\star- \gamma( \hat \nu_{0},\rho_{0},\phi^{(k_0)})-  \alpha_{k_0}(\hat \nu_{0})}{\delta}\rceil+1,
\end{align*}
 without resorting to the high-fidelity model and the original score function $\phi$.
%
%

\subsubsection{Second Statement: Minimal Squared  Coefficient of Variation}

We  present hereafter the proof of the second part of Theorem~\ref{theorem:1}, \ie  which shows that, for a number of samples large enough,  the   squared coefficient of variation   given by the theorem   
is the minimal achievable for all  feasible   $\hat A^{(k_{j-1})}_j$'s.\\

\proof{On the one hand, we show in Remark~\ref{cor:1} of Appendix~\ref{app:remarks}, that  by construction we have the inclusion $A^{(k_{j-1})}_j \supseteq A$.  Moreover, from the almost sure convergence of quantile and error bound approximation (statements {\it i)} and  {\it ii)} of Lemma~\ref{prop:cvMC}), we know  that  $\hat A^{(k_{j-1})}_j$ converges almost surely to $ A^{(k_{j-1})}_j$  as $m \to\infty$. Therefore, we deduce that the inclusion $\hat A^{(k_{j-1})}_j \supseteq A$ holds asymptotically.  This inclusion  is a necessary condition to obtain a finite  variance since we can not guarantee that $ A \setminus \hat A^{(k_{j-1})}_j  \nsubseteq   \textrm{supp}(\mu)$ \footnote{
Indeed, in the case we build a sequence of feasible $\hat A^{(k_{j-1})}_j$'s such that we have $\hat A^{(k_{j-1})}_j \nsupseteq A$ and $ A \setminus \hat A^{(k_{j-1})}_j  \subseteq   \textrm{supp}(\mu)$,  using the assumption  $\nu_{\hat A_j^{(k_{j-1})}}^{\star} \in \mathcal{V}$ we have   the infinite  variance
$
\mathbb{E}[(  p^{IS}_{A,\nu_j}-p_A)^2]= \frac{1}{m_j}(\langle \unit_A \frac{\mu}{\nu_j}, \mu \rangle - p^2_{A})=\frac{1}{m_j}(  p_{\hat A^{(k_{j-1})}_j}\langle\frac{\unit_A}{ \unit_{\hat A^{(k_{j-1})}_j} }  , \mu \rangle - p^2_{A}).
$}.
On the other hand, by  following the reasoning of Appendix~\ref{app:0} where $p_{\hat A^{(k_{j-1})}_j}$ substitutes for $p_{\hat A_j}$,  we  see that assuming  $\nu_{\hat A^{(k_{j-1})}_j}^{\star} \in \mathcal{V}$ and $m_j$ large enough, the  variance writes after the $j$-th iteration as
$
\mathbb{E}[(  p^{IS}_{A,\nu_j}-p_A)^2]
=\frac{p_A}{m_j}( p_{\hat A^{(k_{j-1})}_j}-p_{A}  )$.  
The variance obviously vanishes after the $J$-th iteration as the algorithm sets $\hat A^{(k_{j-1})}_j=A$. 

Let us verify that  this quantity is minimal  among all  feasible   $\hat A^{(k_{j-1})}_j$.  In Remark~\ref{rem:4} of Appendix~\ref{app:remarks}, we show that a necessary condition to obtain a finite variance is that  $ A^{(k_{j-1})}_j \supseteq   A_j $. Since $A_j \supseteq A$,  the  probability $p_{ A^{(k_{j-1})}_j \setminus A}$ is minimal  if the probability $p_{ A^{(k_{j-1})}_j \setminus  A_j}$ is the lowest achievable. 
 For $m_j$ large enough, this is equivalent  to search the minimal  $p_{ \hat A^{(k_{j-1})}_j \setminus  \hat A_j}$ among all  feasible   $ \hat A^{(k_{j-1})}_j$.   
 The result follows since, by  construction  the set $\hat A^{(k_{j-1})}_j $ is the smallest feasible set such that  $\hat A^{(k_{j-1})}_j \supseteq  \hat A_j $, implying that the probability $p_{\hat A^{(k_{j-1})}_j \setminus \hat A_j}$ is the smallest achievable.$\square$

}

\section{Numerical Evaluation:  Pollution Alerts}\label{sec:NumRed}\label{sec:numerical}
We  consider a mass transfer problem  describing the behavior
 of pollutant released by industrial chimneys into the atmosphere, or by a
plant in a river.  The concentration of the pollutant evolves in a compact domain $\Omega \subseteq \Rr^2$. We are interested in the rare event probability that the maximum value of the concentration of the pollutant exceeds a given value in the domain.   

\subsection{Physical Problem}
\begin{figure}[!h]
\begin{center}
\begin{tabular}{ccc}
\includegraphics[width=0.3\textwidth]{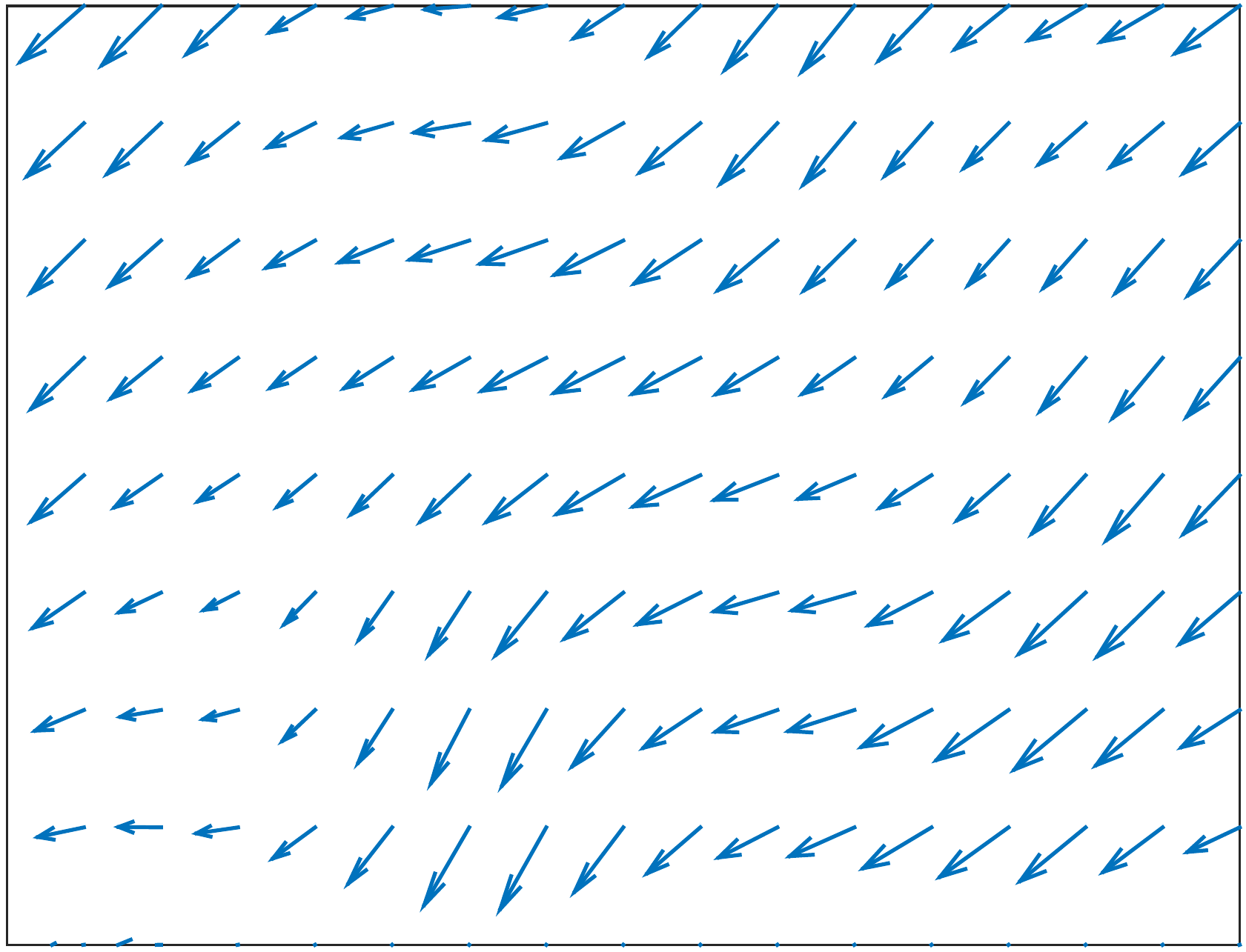} &\includegraphics[width=0.3\textwidth]{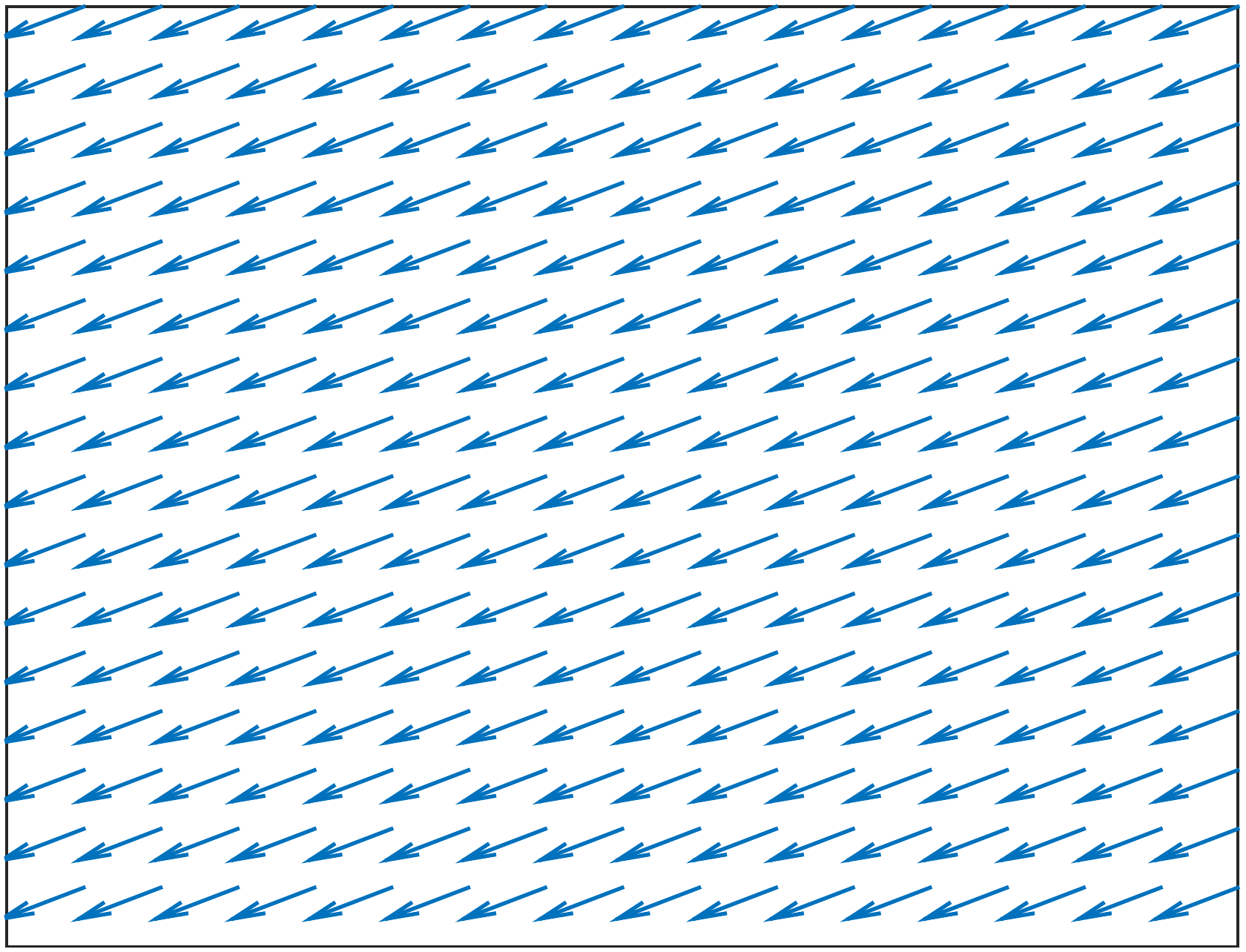}& \includegraphics[width=0.3\textwidth]{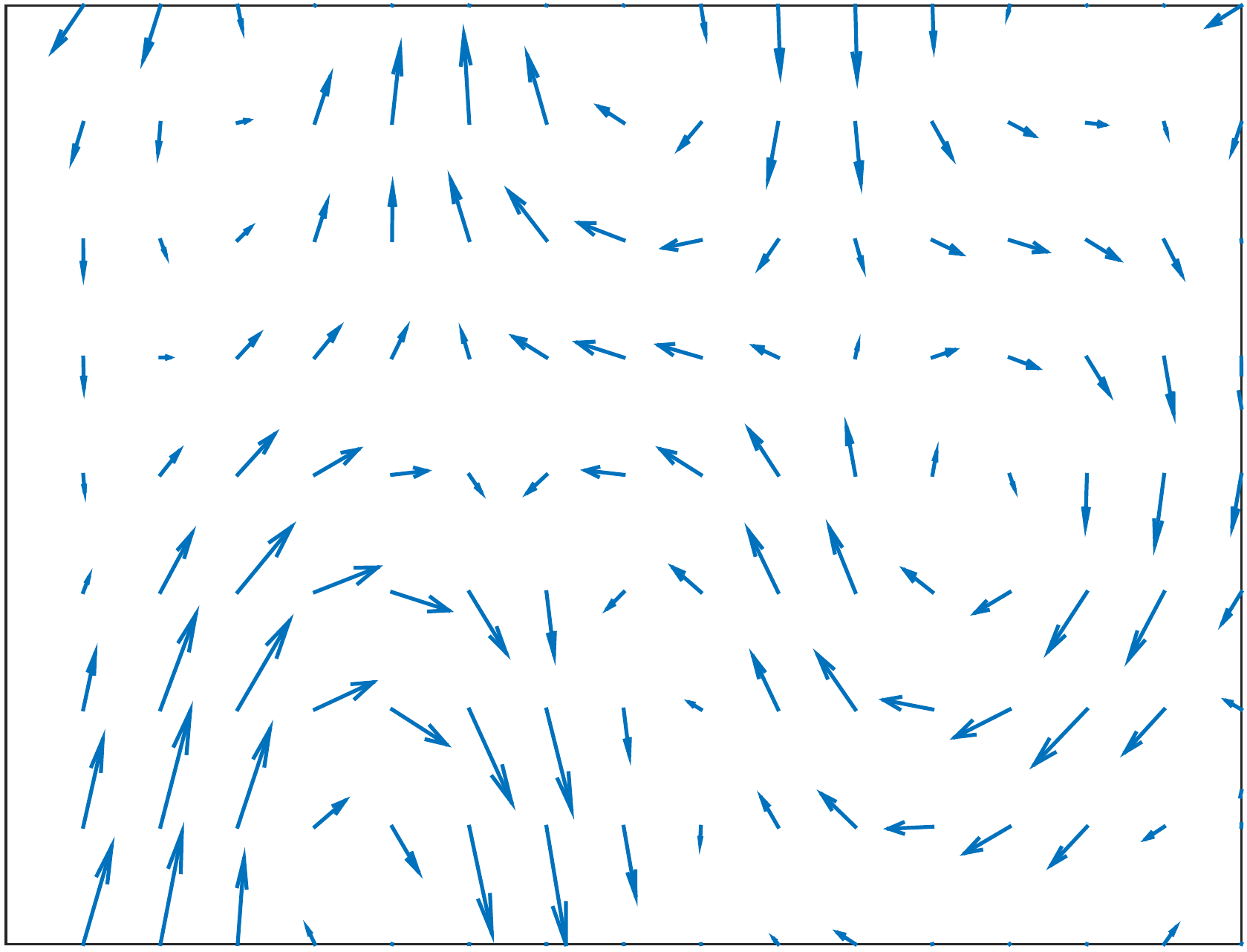} \vspace{-0.25cm}
\end{tabular}
	\caption{{\footnotesize  {  \textbf{Turbulent motion field}. For a sample $x$, motion $b(z,x)$  (left) decomposed  into a constant field (middle) and a divergence-free fBm (right).  \label{fig:6}}}}\vspace{-0.35cm}
		\end{center}
\end{figure}

As detailed in~\cite[Chapter 8.4]{quarteroni2015reduced}, the evolution of the  pollutant 
is modeled by an advection-diffusion-reaction equation, while the emission is
described by a parametrized source term. The pollutant concentration function  $f: \Omega \times \Rr^p \to \Rr^+$ is ruled by the
following  PDE:
\begin{align}\label{eq:ADRproblem}
 \left\{\begin{aligned}
 -\kappa_1\Delta f(z,x)+b(z,x)\cdot \nabla f(z,x)+a_0f(z,x)&={s}(z,x),\quad z\in \Omega\\
 \kappa_1\nabla f(z,x) \cdot n(z)&=0,\quad\quad\quad z\in \partial\Omega,
 \end{aligned}\right.
\end{align}
parametrized by the $p$-dimensional vector $x=(x_1^\intercal,x_2,x_3)^\intercal $ with $x_1\in \Rr^{p-2}$ and $x_2,x_3 \in \Rr$. We considered the domain $\Omega=\begin{bmatrix} 0&1\end{bmatrix} \times \begin{bmatrix} 0&1/2\end{bmatrix} $. 
 The normal to the domain boundary $\partial\Omega$ is denoted by  vector $n(z)$,  
 $\kappa_1=0.03$ represents
the molecular diffusivity  of the chemical species
and  $a_0=0.5$ represents
the intensity of reaction processes. Here, $b(z,x)$ is a turbulent motion field parametrized by $p-2$ coefficients gathered in vector $x_1$. More precisely,  $b(z,x)$ is the (zero-mean) divergence-free fractional Brownian motion (fBm) proposed in~\cite[Proposition 3.1]{heas2014self} of Hurst exponent $H=\frac{1}{3}$, supplied by an additional constant field, see Figure~\ref{fig:6}.  The fBm admits an affine parametrization with $p-3$ real uncorrelated  wavelet coefficients distributed according to the standard normal law, while  the constant field is of unit magnitude and parametrized by a wind direction angle. The number of wavelet coefficients parameterizing the fBm is related to the number of elements considered in the truncated wavelet series, and in turns determines  the motion field resolution. The source
$
s(z,x)=\exp\left( -\frac{(z_2-x_2)^2+(z_3-x_3)^2}{\kappa_2^2}\right),
$
with $z=(z_2,z_3)^\intercal \in \Omega$ describes the pollutant emission, characterized in terms of its position $(x_2,x_3)^\intercal$ and
its spreading $\kappa_2$.  Parameter $x$ will be drawn according to a $p$-dimensional Gaussian random distribution $\mu$. The setup for the Gaussian parameters will be detailed in Section~\ref{sec:numEvalAlgo}.

We consider the weak formulation of~\eqref{eq:ADRproblem} and consider  high-fidelity approximations $f^\star: \Omega \times \Rr^p \to \Rr^+$ of its solutions  via a finite-element method,  see details in \cite[Chapter 2.5]{quarteroni2015reduced}.  

 We are interested in the rare event probability that the maximum value of the concentration of the pollutant over the domain  exceeds a given value $\gamma^\star$. We  thus define the probability $p_A$ with $A=\{x\in \Rr^p : \phi(x) \ge \gamma^\star\}$, where the score function is defined as the sup norm, \ie
$
\phi(x)=\| f^\star(\cdot, x) \|_\infty .
$

\subsection{Reduced Basis Approximations}

\begin{figure}[!h]
\begin{center}
\begin{tabular}{ccc}
\hspace{-0.5cm}\includegraphics[width=0.4\textwidth]{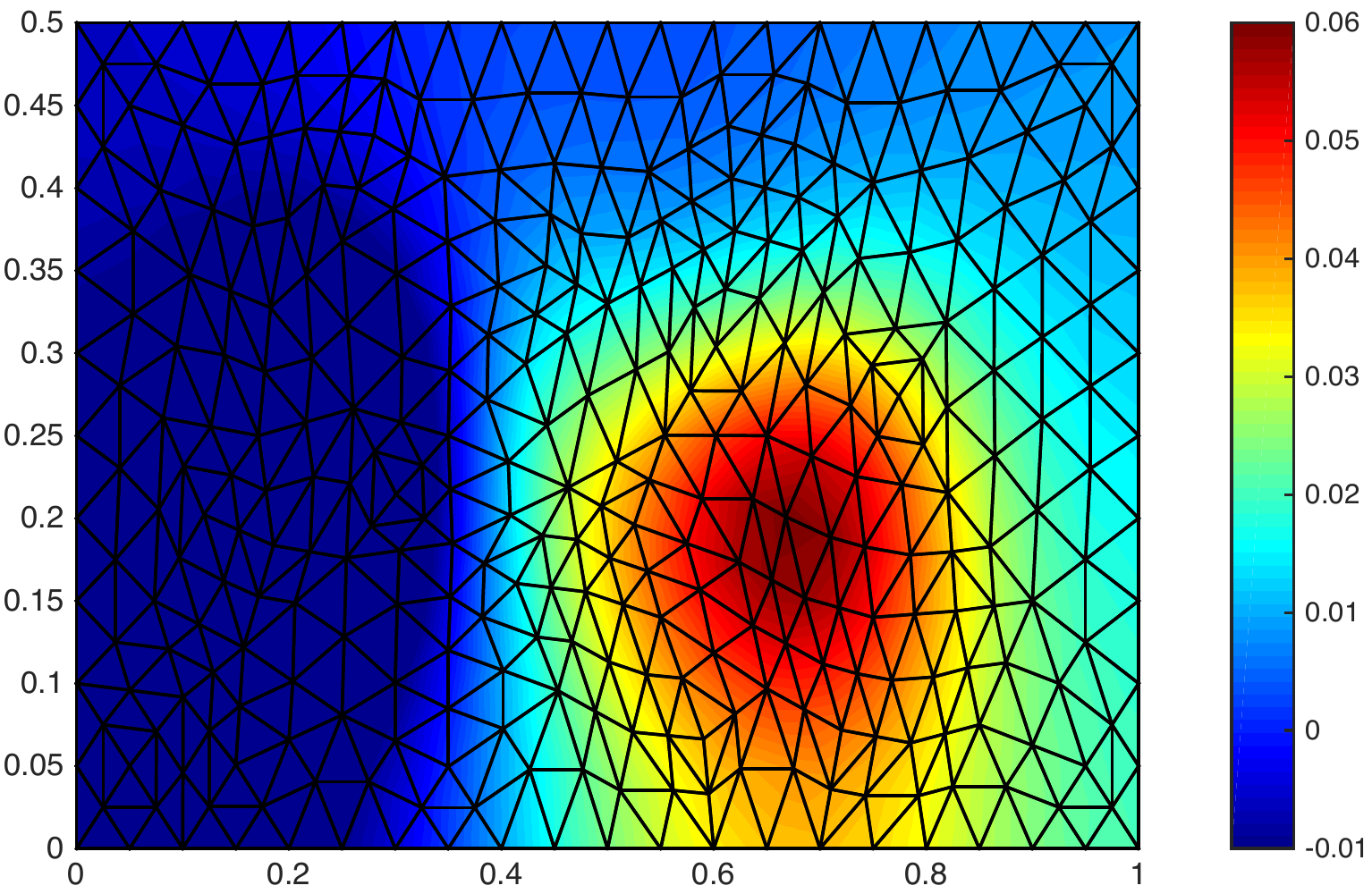}&\hspace{-1.25cm}\includegraphics[width=0.4\textwidth]{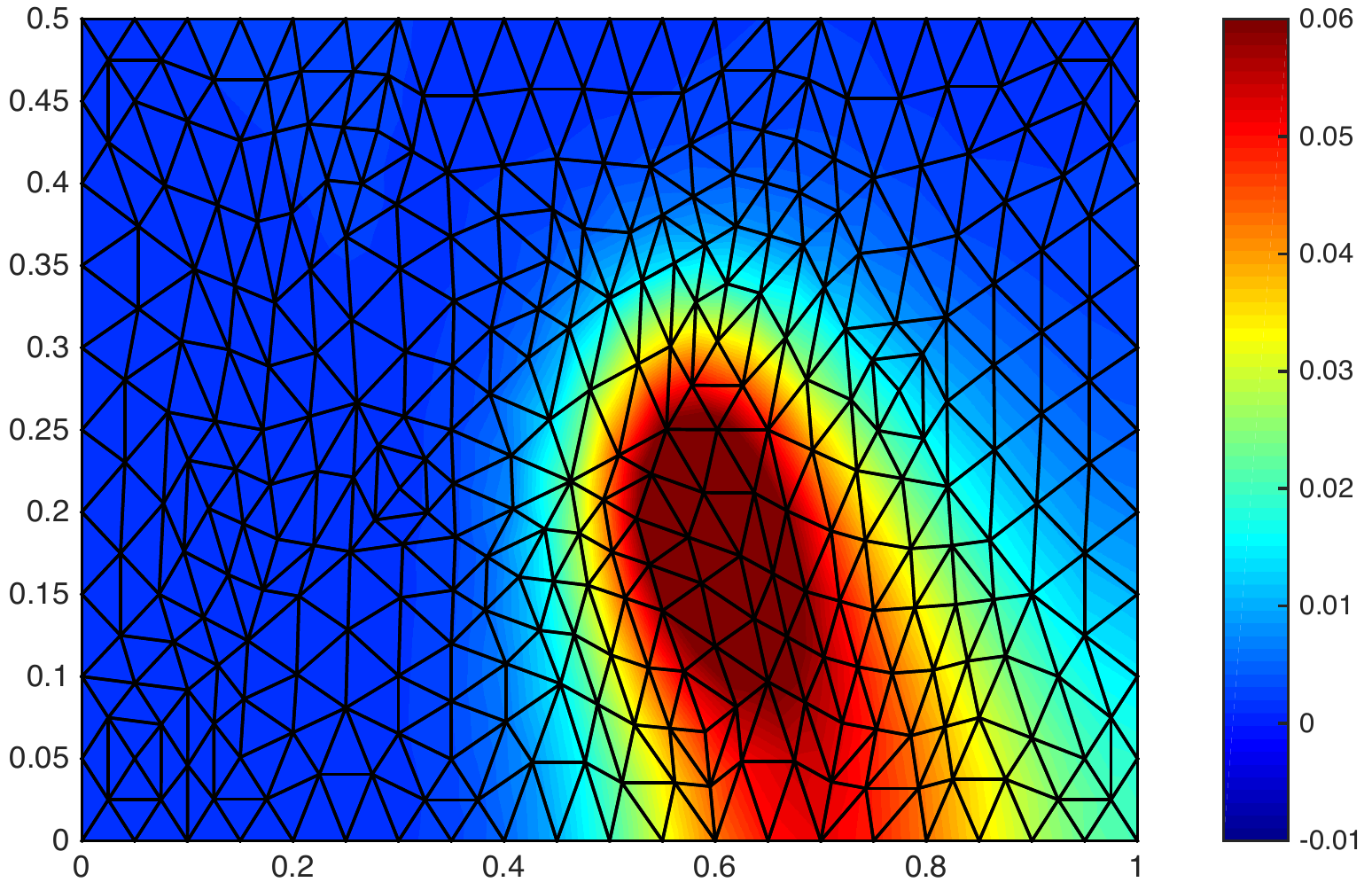}&\hspace{-1.25cm}\includegraphics[width=0.4\textwidth]{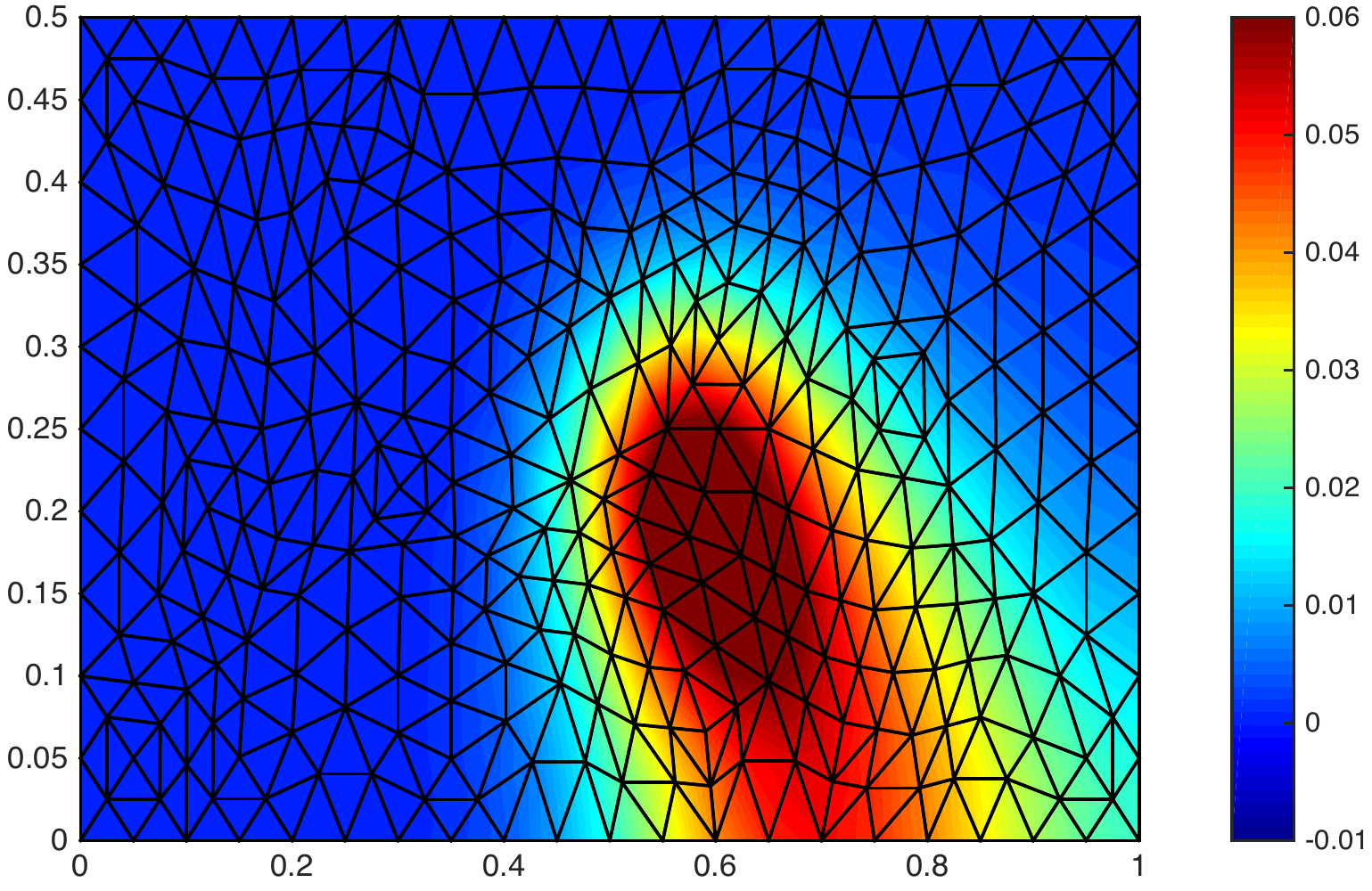}\\
\hspace{-1.cm} $d_k$=5  &\hspace{-2.5cm} $d_k$=50  & \hspace{-1.5cm}High-fidelity ($q=648$)  \vspace{-0.25cm}
\end{tabular}
	\caption{{\footnotesize  {  \textbf{Approximation of  pollutant concentration}. For a sample $x$, mesh and solution  by Galerkin projections $f^{(k)}(z,x)$ or by the high-fidelity model $f(z,x)$ (with spreading $\kappa_2=0.15$).  \label{fig:1}}}}\vspace{-0.2cm}
		\end{center}
\end{figure}

We use principal orthogonal decomposition (POD) to compute the set of nested approximation sub-spaces  $\{\Sps_\idxss\}_{\idxss=1}^{K}$, where $d_K=250$. We  use $2000$ representative solutions, so-called snapshots.  Since $s(z,x)$ is nonlinear in $x$, problem~\eqref{eq:ADRproblem} does not fulfill the assumption
of affine parametric dependence necessary to take advantage of the off-line/on-line decomposition. In consequence, we set up  an approximate affine expansion by means of empirical interpolation
method using $30$ interpolation points. In this manner, we generate for any $x$ and any $k$ the $d_k$-dimensional RB approximation $f^{(k)}(z, x)$ of
the high fidelity solution $f^\star(z,x)$, see details in \cite[Chapter 10.5]{quarteroni2015reduced}. Examples of reduced basis approximations are displayed in Figure~\ref{fig:1}.  We fix a hierarchy of reduced model $f^{(k)} \in \mathcal{F}$ related to the set of dimensions $\mathcal{K}$. The different choices for the set $ \mathcal{K}$ will be detailed in Section~\ref{sec:numEvalAlgo}. 
The  a posteriori error estimate $\Spwidth_\idxss(x)$ at point $x$ for the reduced model $f^{(k)}$  is computed  using an  off-line/on-line decomposition~\cite{quarteroni2015reduced}.

Accordingly we derive for  $k=1,\cdots,K$, the score function approximation from the RB approximation $f^{(k)}$ as
$
\phi^{(k)}(x)=\| f^{(k)}(\cdot, x) \|_\infty.
$
Using~\eqref{eq:alphaHat}, this leads  to bound the approximation error norm as
$
 \alpha_k(\hat \nu)= \max_{i=1,\cdots,m}\epsilon_{k}(z_i),
$
where  $z_1,\cdots,z_{m}$ are  i.i.d.  samples from $\nu \in \mathcal{V}$.

\subsection{Parametrization of the CE methods}\label{sec:paramGauss}

The set  $\mathcal{V}$ is chosen to be the family of $p$-dimensional normal distributions
$$
\nu(\theta)=\frac{1}{\sqrt{|2\pi \Sigma_\nu|}}\exp\left(-\frac{1}{2}(x-m_\nu)^\intercal \Sigma_\nu^{-1}(x-m_\nu)\right)
$$
 parametrized by vector  $\theta \in \Rr^{p(p+1)}$, gathering the components of the mean $m_\nu \in \Rr^p$ and of the covariance matrix $\Sigma_\nu  \in \{M \in \Rr^{p \times p}: |M|  \ge 0,  M^\intercal=M\}$. 
 On the one hand, we notice in this example that, unfortunately, the zero-variance biasing distribution $\nu_A^{\star}$ defined in~\eqref{eq:optDensity} does not belong to the Gaussian family $\mathcal{V}$. Therefore, we expect the optimal distribution $\nu_A^{\theta^\star} \in \mathcal{V}$ to have a non-zero variance. 
 On the other hand, this family presents the advantage to yield a closed-form solution $\nu_j$ to problem~\eqref{eq:minDiv4_}. Indeed, cancelling the gradient of the objective function of~\eqref{eq:minDiv4_}, we can show that the parameters $(m_{\nu_j},\Sigma_{\nu_j})$ of the Gaussian distribution $\nu_j$ are solution of  linear systems. More precisely, we find that  $m_{\nu_j}$ is  the solution of \vspace{-0.15cm}
 \begin{align*}  
\sum_{i=1}^{m_{j-1}} \unit_{\hat A^{(k_{j-1})}_j}(z_i) \frac{\mu(z_i)}{\nu_{j-1}(z_i)} (z_i-m_{\nu_j})&=0,
\end{align*} 
while $\Sigma_{\nu_j}$ solves\vspace{-0.15cm}
 \begin{align}\label{eq:solCovGauss}  
\sum_{i=1}^{m_{j-1}} \unit_{\hat A^{(k_{j-1})}_j}(z_i) \frac{\mu(z_i)}{\nu_{j-1}(z_i)} (I_p- (z_i-m_{\nu_j})(z_i-m_{\nu_j})^\intercal\Sigma_{\nu_j}^{-1})&=0,
\end{align} 
where $I_p$ is the p-dimensional identity matrix. We verify that  the latter  solution is  feasible by construction, \ie symmetric and positive semi-definite. Nevertheless, these systems are full-rank (implying a unique solution) only if sufficient $z_i$'s (distributed according to $\nu_j$) belong to the set ${\hat A^{(k_{j-1})}_j}$. Hopefully,  it is straightforward to show that this happens  for a sample size $m_{j-1}$ large enough.

 \subsection{Experimental  Setup}\label{sec:numEvalAlgo}
 We detail  in this section practical considerations concerning  our numerical simulations. 
 
 Distribution $\mu$ is  chosen to be an uncorrelated $p$-dimensional  Gaussian  of mean $\begin{pmatrix} 0.8&0.15&17\pi/18& 0 &\cdots &0\end{pmatrix}^\intercal$ and characterized by an  identity covariance matrix.  We design the following  experimental benchmark.
 
\begin{center}
\begin{tabular}{c|c|c|c|c|c}
 Experiment  &$q$&$p$&$\kappa_2$&$\gamma^\star$&$ p_A$  \\
\hline
 \#1   &648 &$3$ &0.25 &1.76& $1.3\times 10^{-6}$\\
 \#2   &648 &$3$ &0.25 &1.8 & $1.2 \times 10^{-10}$\\
 \#3  &$17024$ &${13}$&0.15 &1.62 &$2.4\times 10^{-5}$\\
 \#4  &$17024$ &${21}$&0.15 &1.62 &$2.1\times 10^{-5}$\\
\end{tabular}
\end{center}

 The solution of the discretized system, the RB approximations and the a posteriori error bounds are computed via  the Matlab\textsuperscript{\textregistered} toolbox ``redbKIT'' available at \url{http://redbkit.github.io/redbKIT}. We accelerate substantially  a posteriori error estimation  by rewriting the procedure with matrix products optimized for Matlab. 

The rare event probabilities are estimated using:  \vspace{-0.2cm} \\
 
 \begin{itemize}
 \item the {\bf standard} CE method (Algorithm~\ref{algo:1}) described in \cite{Homem-de-Mello02rareevent},
 \item the {\bf  pre-conditioned} CE method (Algorithm~\ref{algo:3}) introduced in \cite{peherstorfer2017multifidelity},
 \item the {\bf  proposed}  CE method (Algorithm~\ref{algo:2}).\vspace{-0.2cm}  \\
 \end{itemize}

For each of thes experiments, in order to obtain (an approximation of) $p_A$ displayed in the previous table, we first compute reference probabilities by averaging $ p_{A,\nu}^{IS}$ over  a set of $30$ estimates  for  experiment~\#1 and ~\#2 (resp. 10 estimates for experiment~\#3 and ~\#4), where each of the rare event probability estimates was obtained using the standard CE and a number of samples of $m=10^4$.
We consider a  set of specific  values of $m\in  \mathcal{M} =\{10^3,2\times10^3,4\times10^3,6\times10^3,8\times10^3,10^4\}$ and  the set  $\mathcal{F}$ of  models of  dimension in  the set $\mathcal{K}$. In particular,  for experiment~\#1 and \#2 we identify $\mathcal{K}$ either to $\mathcal{K}_2=\{150,q\}$, $\mathcal{K}_6= \{50,100,150,200,250,q\}$ or $\mathcal{K}_{21}=\{50,60,\ldots,250,q\}$, while for   experiment~\#3 and ~\#4   we identify $\mathcal{K}$ either to $\mathcal{K}_2=\{100,q\}$, $\mathcal{K}_5= \{100,125,150,175,q\}$ or $\mathcal{K}_8=\{50,75,100,125,150,175,200,q\} $.

 We finally provide some additional details on the  parametrization of the CE methods. The initial quantile parameter is set to $\rho=0.2$ and the minimal step size is chosen to be $\delta=10^{-2}$.   We avoid  the general tendency of densities ratio in the IS estimate to go to infinity for large values of $p$~\cite{au2003important}, 
by setting the eigenvalues  of the covariance matrix solving~\eqref{eq:solCovGauss}  to a minimal value of $5.10^{-5}$ for experiments~\#1,~\#2 and~\#3 and of $5.10^{-4}$ for experiment~\#4. Note that a similar strategy is adopted in~\cite{peherstorfer2017multifidelity}. 

Besides,  we mention that in the context of the  proposed CE method, we substituted $\alpha_{k_{j-1}}(\hat \nu_{j-1})$ by $\alpha_{k_{j}}(\hat \nu_{j})$ in $\tilde \gamma$ appearing in \eqref{eq:deltaGamma} to avoid unreachable values of $\tilde \gamma$ with the reduced model $f^{(k_j)}$ in the case the error comited with $f^{(k_{j-1})}$ is too large in the first levels of the hierarchy. In  the worst-case scenario, this substitution will yield $2(\alpha_{k_{j-1}}(\hat \nu_{j-1})-\alpha_{k_{j}}(\hat \nu_{j}))/\delta$ supplementary iterations in the proposed CE method.

 \subsection{Results}

 The quality and efficiency of rare event estimation is evaluated  according to the following criteria.
\begin{itemize}
\item {\bf Variance}. The estimator SCV defined in~\eqref{eq:SCV} was approximated by the empirical average over  $10$ estimates  of 
${( p_{A,\nu}^{IS}-p_A)^2}/{ p_A^2}$. Here $p_A$ denotes the  reference probability  computed  previously and $ p_{A,\nu}^{IS}$ denotes the current rare event probability estimate.
\item  {\bf Runtime}. This measurement is obtained by  averaging  the algorithm's total runtime  in seconds over  of a set of $10$ rare event estimation processes. We use the  4 cores of  a 2,8 GHz Intel Core i7 processor with 16 GB RAM running a Matlab\textsuperscript{\textregistered} parallelized implementation. 
\item {\bf Number of iterations}.  The number of iterations at each level  of the CE algorithms   was  averaged over a set of $10$ rare event estimation processes.\\
\end{itemize}

 \begin{figure}[!h]
\vspace{-0.35cm}\begin{center}
\begin{tabular}{cc}
\hspace{-0.25cm}experiment~\#1 &\hspace{-0.5cm} experiment~\#3 \\
\hline
\hspace{-0.25cm}\includegraphics[height=0.4\textwidth]{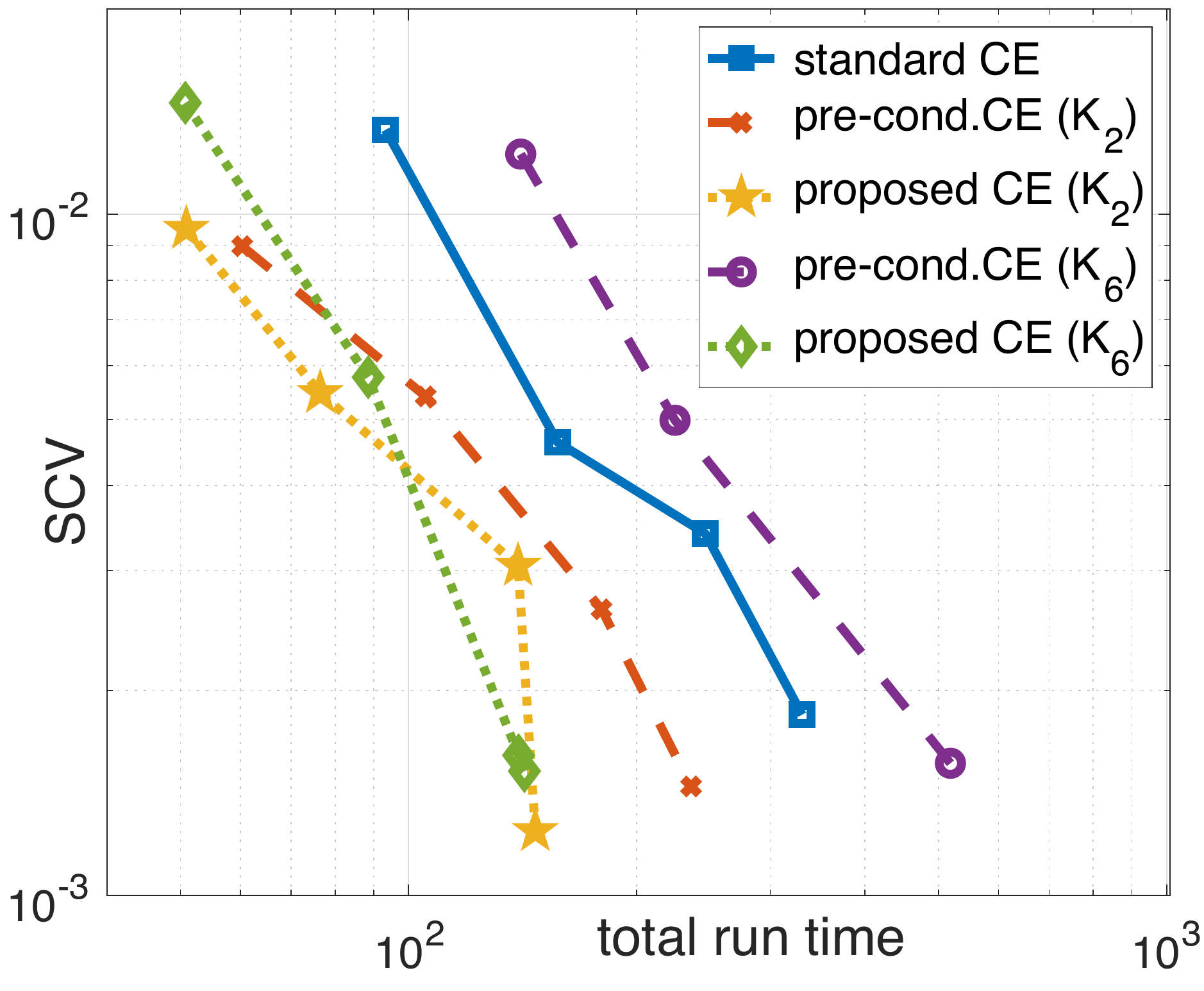}&\hspace{-0.5cm} \includegraphics[height=0.4\textwidth]{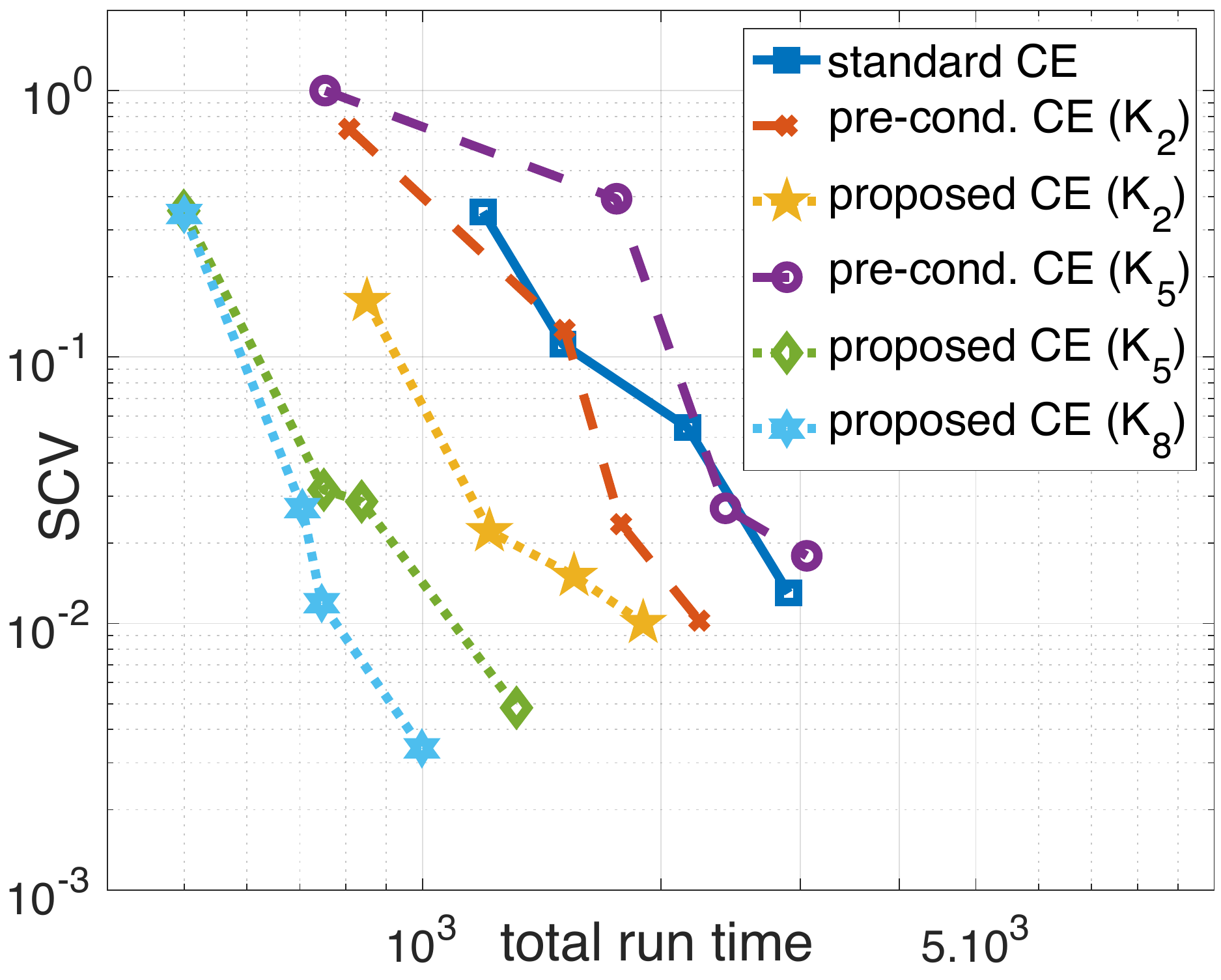}\vspace{-0.35cm}\\

\vspace{-0.35cm}
\end{tabular}
	\caption{{\footnotesize  {  \textbf{Evaluation in terms of variance and runtime}.  Plots are obtained  varying $m\in \mathcal{M}$ and $\mathcal{K} \in \{\mathcal{K}_2,\mathcal{K}_6\}$ (experiment \#1) or $\mathcal{K} \in \{\mathcal{K}_2,\mathcal{K}_5,\mathcal{K}_8\}$ (experiment \#3). \label{fig:2}}}}\vspace{-0.2cm}
		\end{center}
\end{figure}

\begin{figure}[!h]
\begin{center}
\begin{tabular}{ccc}
\multicolumn{2}{c}{experiment~\#1}&experiment~\#2\\
\hline
\hspace{-1.5cm}\includegraphics[height=0.32\textwidth]{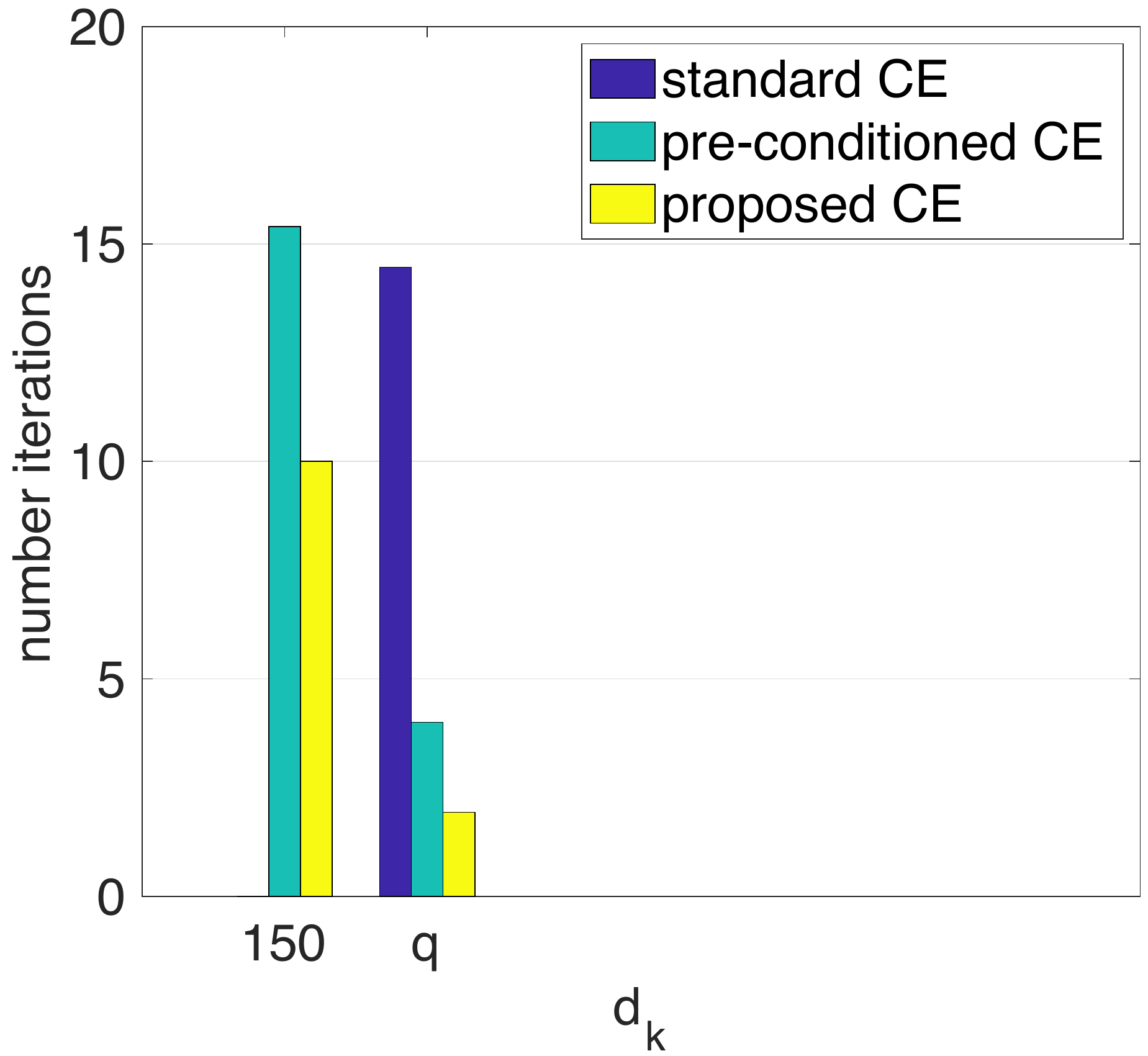}&\includegraphics[height=0.32\textwidth]{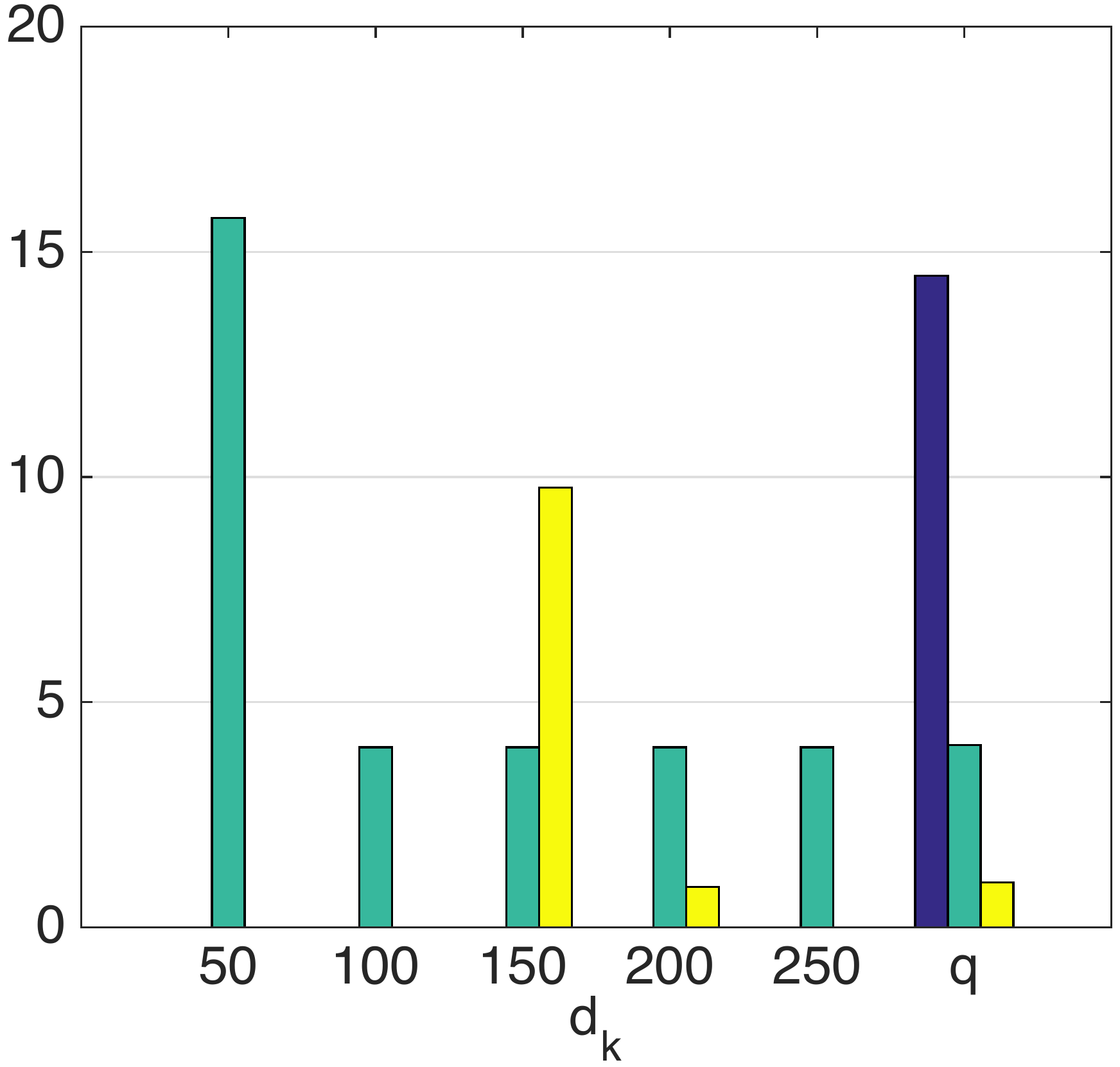}
&\includegraphics[height=0.32\textwidth]{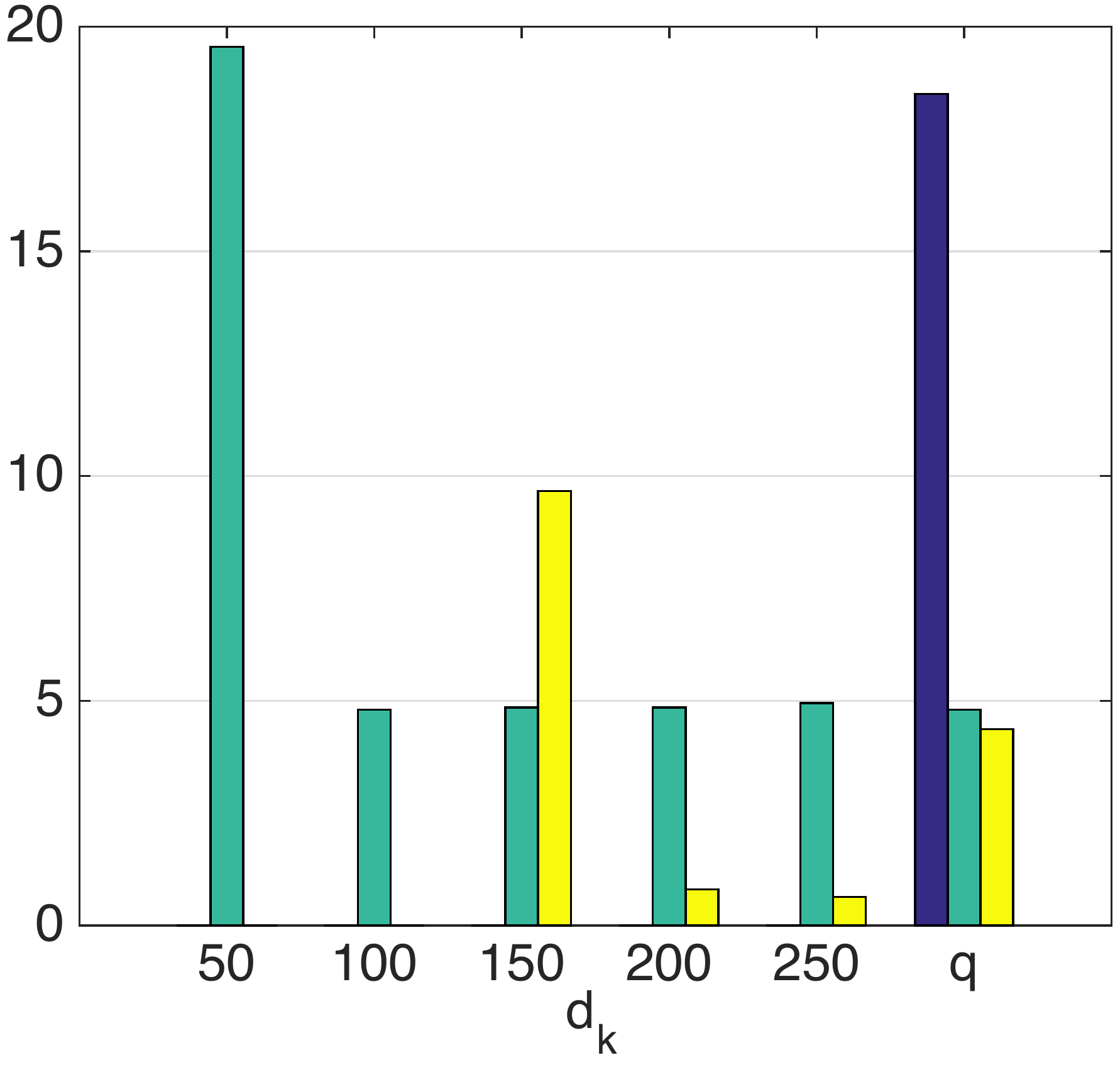}\\
\multicolumn{2}{c}{experiment~\#3}&experiment~\#4\\
\hline
\hspace{-1.5cm}\includegraphics[height=0.32\textwidth]{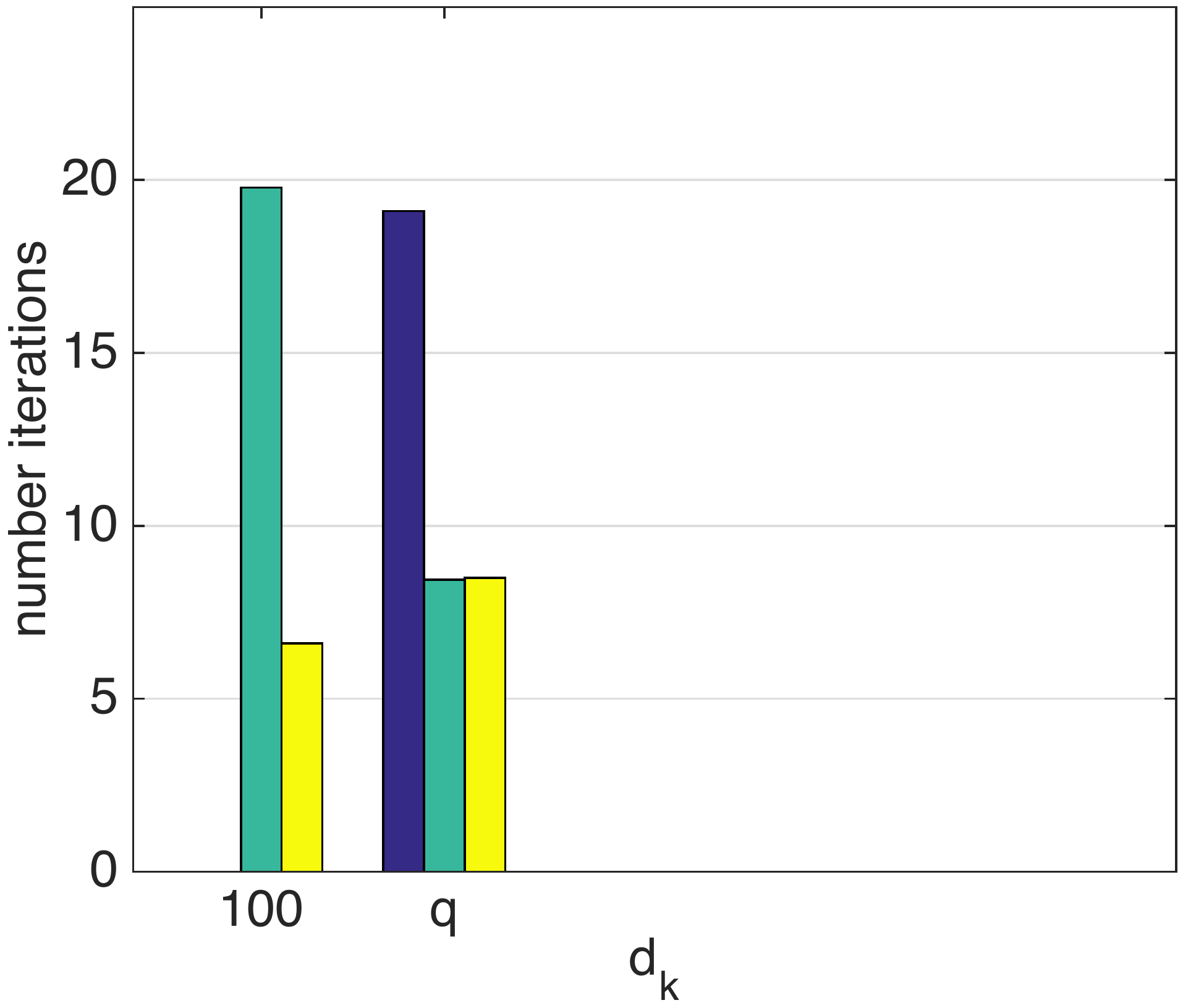}&\includegraphics[height=0.32\textwidth]{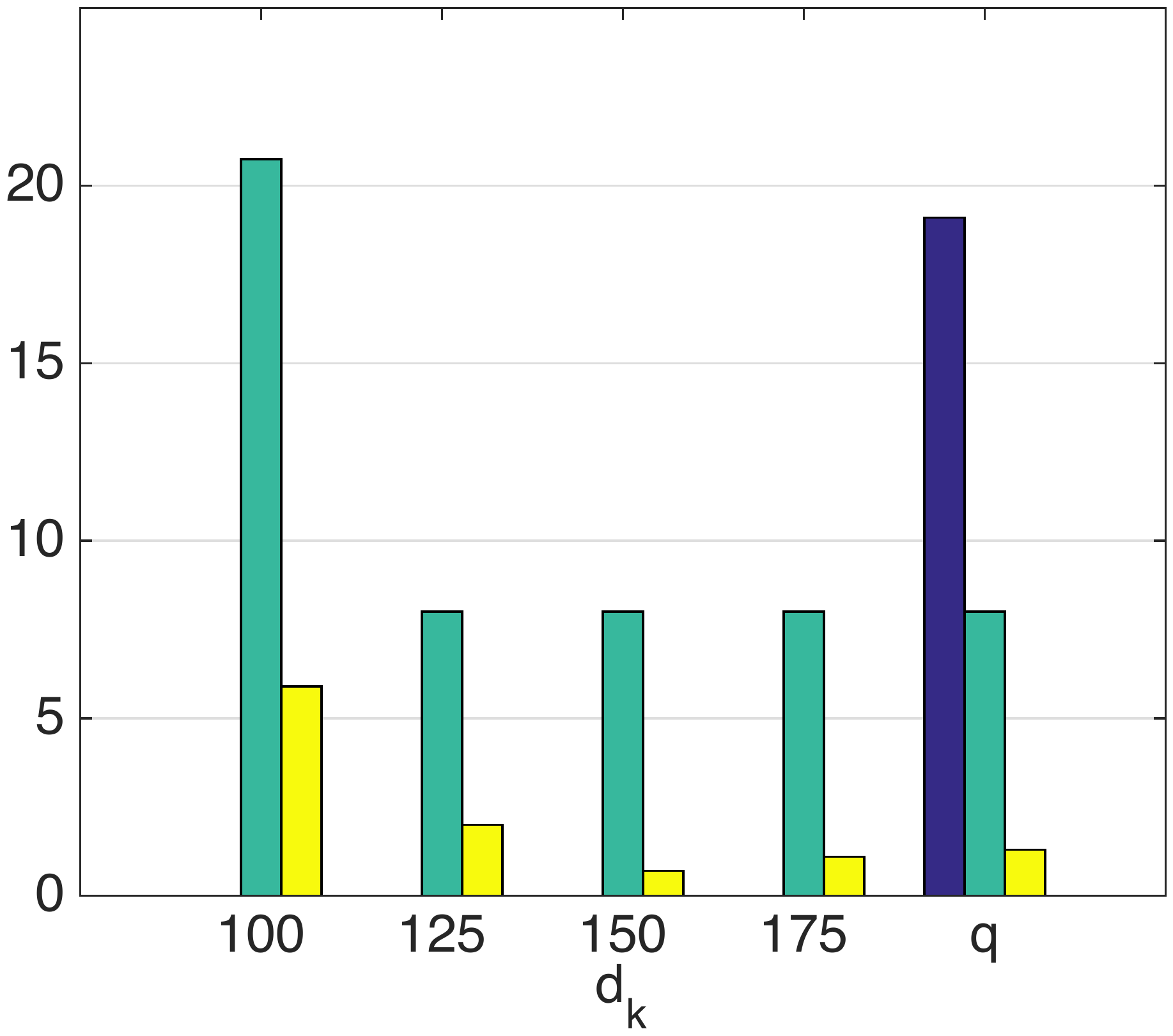}&\includegraphics[height=0.32\textwidth]{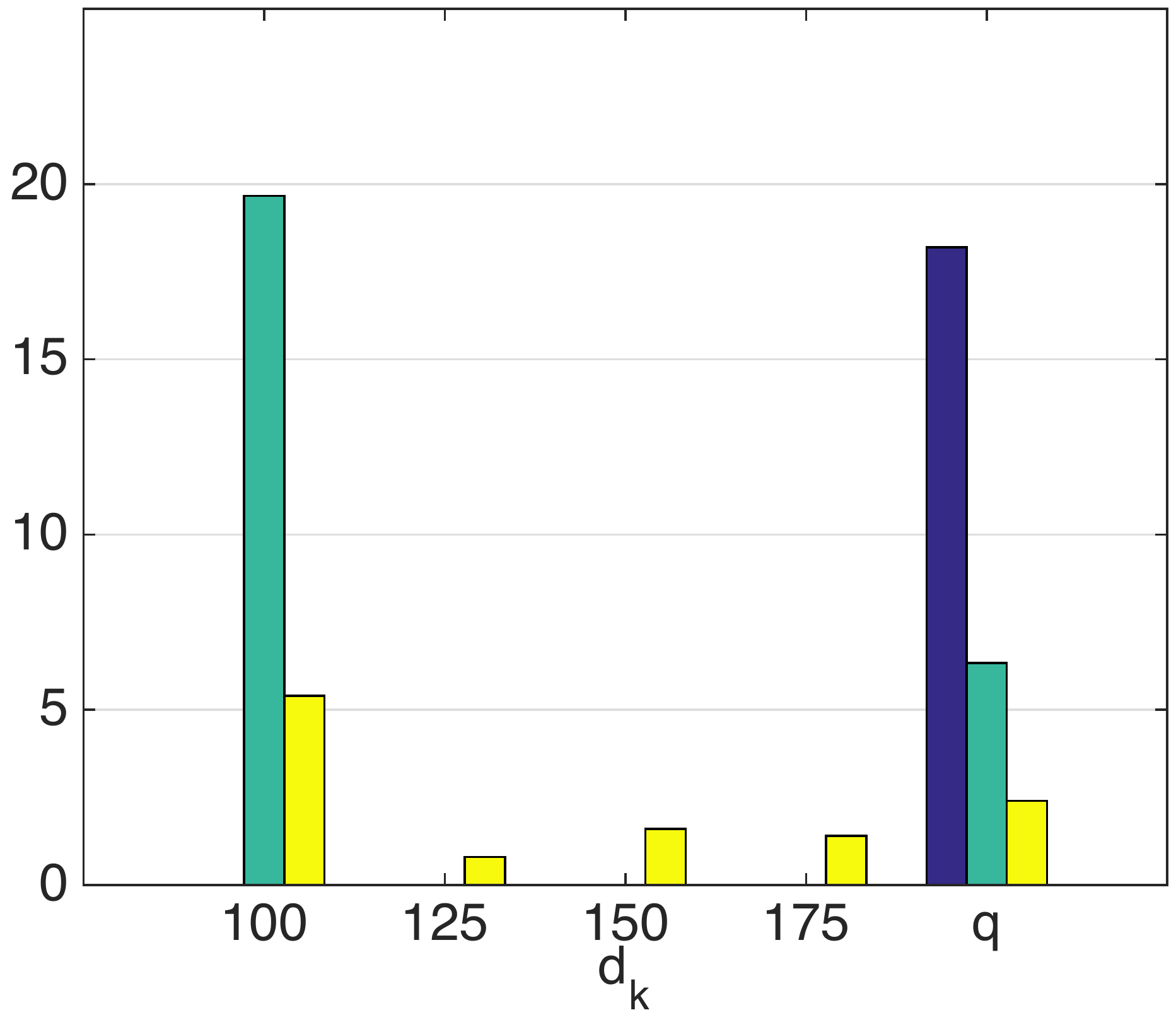}    \vspace{-0.45cm}
\end{tabular}
	\caption{{\footnotesize  {  \textbf{Evaluation in terms of iteration number} Average iteration number for the different algorithms  at each level of the hierarchy of reduced models  of dimension $d_k$ in the set $\mathcal{K}_2$, $\mathcal{K}_5$ or $\mathcal{K}_6$ (The most favorable hierarchy was chosen in experiment~\#4, \ie $\mathcal{K}_2$ and $\mathcal{K}_5$ respectively for the pre-conditioned and the proposed methods).  \label{fig:3}}}}\vspace{-0.5cm}
		\end{center}
\end{figure}

We show in Figure~\ref{fig:2} the performance of the  standard, pre-conditioned and proposed  algorithms in terms of the estimator's SCV as a function of the overall algorithm runtime. Plots are obtained by varying the sample size $m\in \mathcal{M}$ and $\mathcal{K} \in \{\mathcal{K}_2,\mathcal{K}_6\}$ for experiment \#1 or $\mathcal{K} \in \{\mathcal{K}_2,\mathcal{K}_5,\mathcal{K}_8\}$ for experiment \#3. 
The bar-plots of Figure~\ref{fig:3} display the average number of iterations spent by the 3 different algorithms in each level of the reduced model hierarchies in the case $m=10^4$ for  experiments~\#1 and~\#2, $m=8.10^3$ for  experiment~\#3 and  $m=6.10^3$ for  experiment~\#4 .

We observe in Figure~\ref{fig:2}  that, in the case of experiment~\#1, the use of a hierarchy of reduced models can lower significantly the algorithm runtime while yielding a similar SCV for the IS estimator. The magnitude of the gain reaches nearly   half a decade, although the problem is characterized by a relatively moderate ratio between the dimension of the high-fidelity model $q=648$ and the dimensions of the reduced models.  The pre-conditioned and the proposed methods both achieve similar performances  in the case of a 2-level hierarchy. However, in the case of a 6-level hierarchy, although the proposed CE method still brings a significant gain, we remark that the pre-conditioned CE method fails in reducing (and even increases) the algorithm runtime.  This  illustrates the fact that selecting the reduced model accuracy in the CE method is mandatory to guarantee a runtime reduction in general. 
The bar-plots of Figure~\ref{fig:3} show clearly that  for  $\mathcal{K}_6$, some levels  of the hierarchy used by the pre-conditioned algorithm are  ignored by our  method, and furthermore, that our method tends to perform  fewer iterations on the selected levels of the hierarchy. According to Figure~\ref{fig:2}, the accuracy of the two methods is similar. We conclude that, on the one hand, most of the levels used by the pre-conditioned method are  useless and, on the other hand, it is unnecessary  to iterate until convergence on each level of the hierarchy. 
 

In experiment~\#2, we consider the lower probability in order of $10^{-10}$ (instead of $10^{-6}$ for experiment~\#1). Interestingly, in order to characterize this very rare event,  the bar-plots of Figure~\ref{fig:3} indicate  that the proposed method  with $\mathcal{K}=\mathcal{K}_6$ performs identically as for experiment~\#1 on the first levels of the hierarchy (for $d_k=50,100,150$ and $200$), and adds only a few iteration on the last levels (for $d_k=250$ and $648$).  This behavior contrasts with the pre-conditioned method which spoils several iterations at each level of the hierarchy to reach convergence. Moreover, we mention that, for the proposed method, the run time acceleration is in order of $\times 2$  for experiment~\#1 and~\#2 (the SCV preserves the same order of magnitude). This is not the case for  the pre-conditioned method, which slows down the estimation by a factor of about $2/3$ for   experiment~\#1, while an acceleration of about $3/2$ is observed for experiment~\#2.

\begin{figure}[!h]
\begin{center} \vspace{-0.45cm}
\begin{tabular}{cc}
\multicolumn{2}{c}{experiment~\#1}\\
\hline
\includegraphics[height=0.32\textwidth]{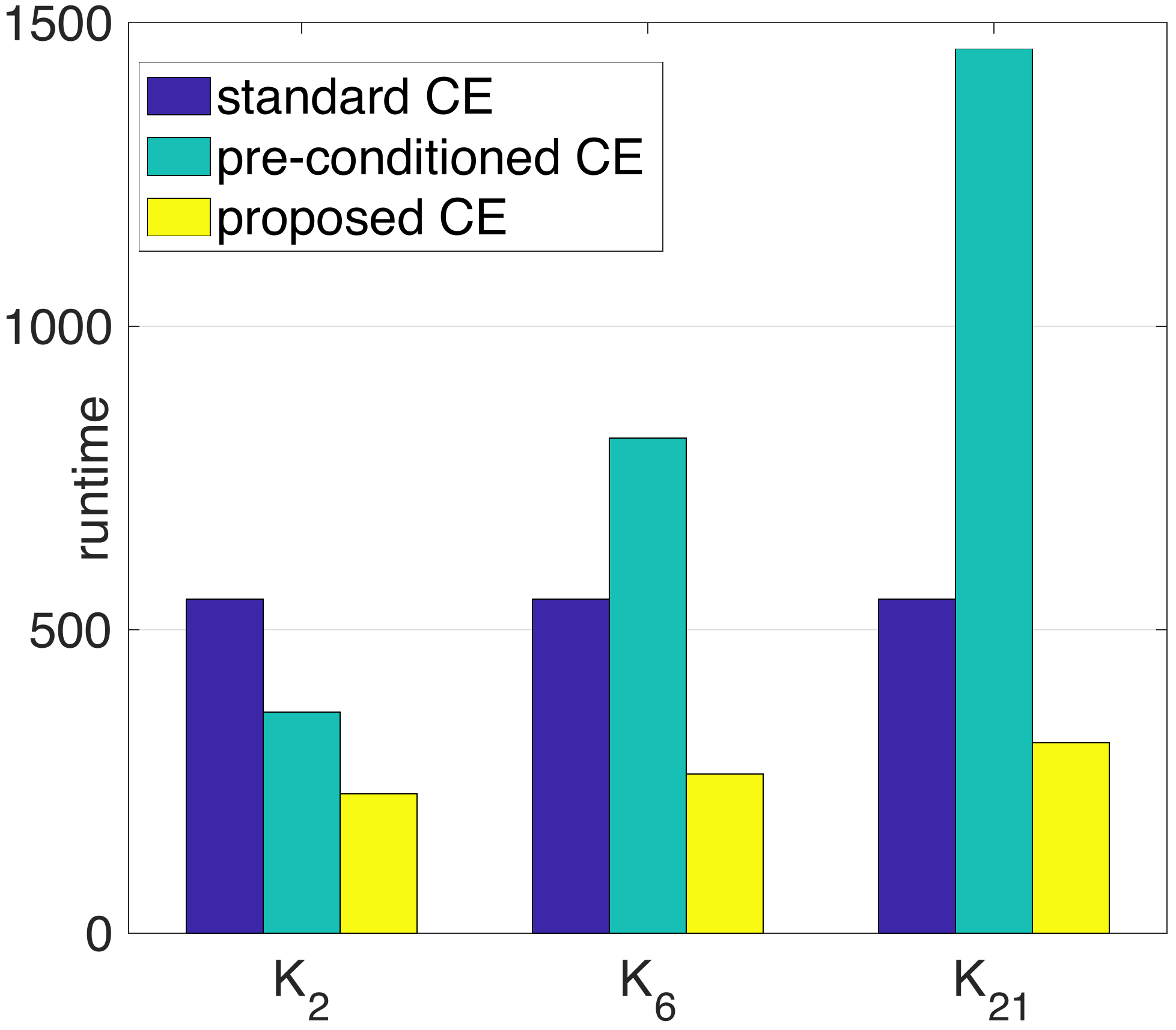}
&\includegraphics[height=0.32\textwidth]{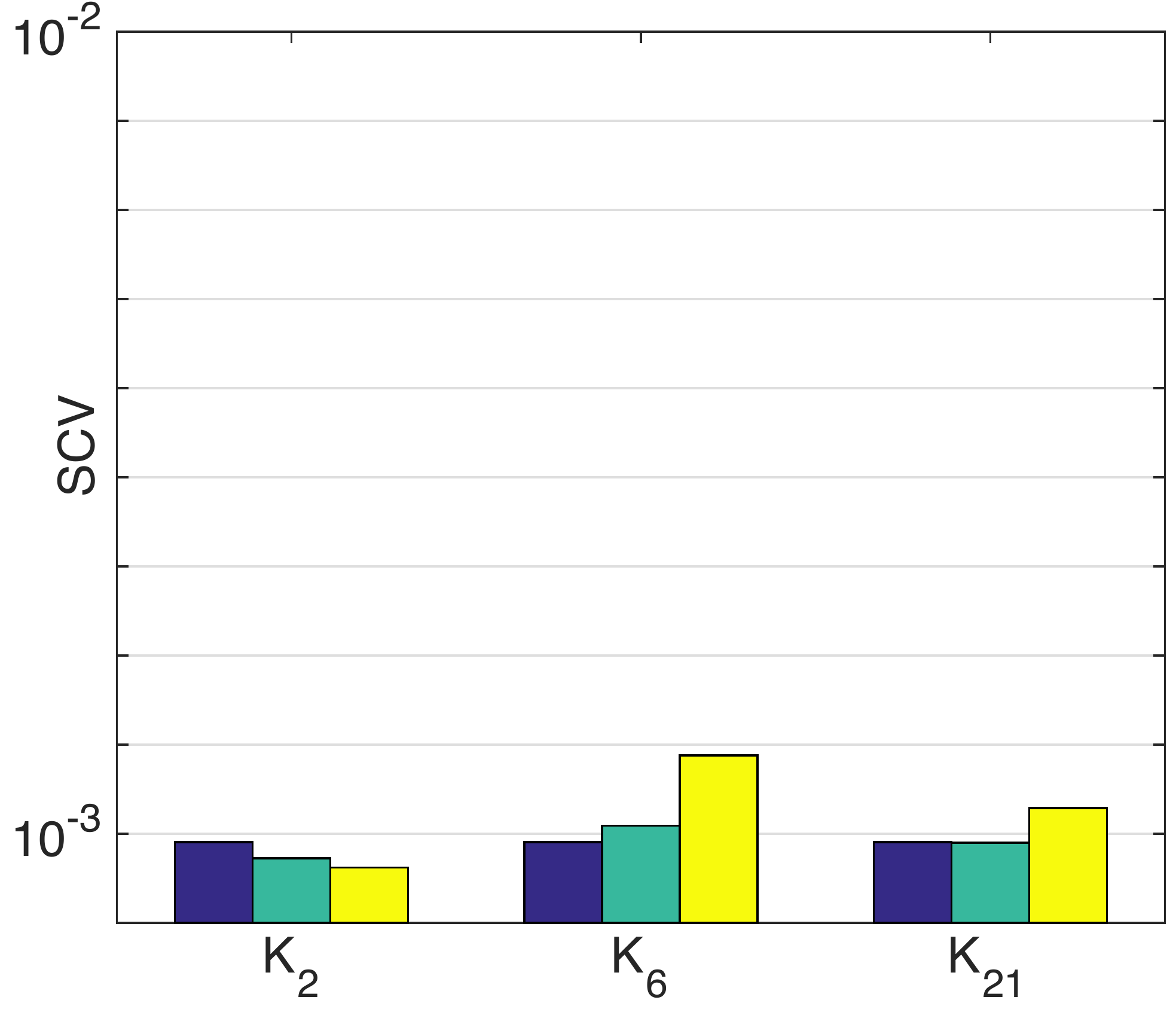}
\vspace{-0.35cm}
\end{tabular}
	\caption{{\footnotesize  {  \textbf{Influence of the number of levels of the hierarchy}.  Results obtained with  the different algorithms for a 2-, 6- or 21-level hierarchy denoted respectively  $  \mathcal{K}_2, \mathcal{K}_6$ or ${\mathcal{K}_{21}}$ for $m=10^4$.    \label{fig:4}}}}\vspace{-0.45cm}
		\end{center}
\end{figure}

To push  our analysis further, we investigate  the influence    of the number of levels of the hierarchy on runtime and SCV. The bar-plots  of Figure~\ref{fig:4} show the influence related to experiments \#1  for a sample size of $m=10^4$ with  $\mathcal{K}\in \{ \mathcal{K}_2,\mathcal{K}_6, {\mathcal{K}_{21}}\} $.   We remark that the runtime of our  method is stable as the number of levels in the hierarchy increase. The behavior of the pre-conditioned method is exactly at the opposite: runtime increases significantly with the number of levels. Nevertheless, this runtime stability or increase does  not  impact the quality of the estimation in experiments \#1 as shown by the stable behavior of the SCV.

Plots of Figure~\ref{fig:2} related to experiment~\#3,   show the behavior of the algorithms when we increase by more than a factor $10$ the  dimensionality of both, the random field parameter and the simulation. More precisely, we substitute in model \eqref{eq:ADRproblem} a constant advection field ($p=3$) for a divergence-free fBm ($p=13$) and we increase the number of finite elements from $q=648$ to $q=17024$. We still observe in this high-dimensional setting the good performance of the proposed method. The plot indicates that a  hierarchy with at least 5 levels   accelerates the rare event estimation process by a factor 3. Using only $2$ levels in the hierarchy slightly reduces the gain brought by the proposed method. The comparable performances obtained for a $5$- or $8$-level hierarchy  confirm the weak sensitivity of the method to the number of levels.   In contrast, the pre-conditioned CE algorithm fails to accelerate the estimation process. In the best scenario, a $2$-level hierarchy only succeeds to reach the performance of the standard CE method. As attested by the bar-plots in  Figure~\ref{fig:3}, these  poor results are induced by the waste of iterations used to reach convergence at each level of the hierarchy, which in turns drives away the biasing density from the optimal one.  On the contrary, the proposed method tends to distribute and spare the iterations among the different levels of the hierarchy.  

As shown in Figure~\ref{fig:2},  the behavior of the three different algorithms remains similar  for experiment~\#4 in terms of the number of iterations. However,  the performances  deteriorate significantly in terms of variance: the SCV of the proposed or the standard methods increase by a factor 4 while the one related to the pre-conditioned method explodes by a factor 50. This deterioration occurring when the dimension of the parameter space is increased up to $p=21$ is likely to be the effect of the curse of dimensionality prevailing for IS~\cite{au2003important}. Nevertheless, the SCV of the proposed method remains about 4 time smaller than the SCV of the standard method,  suggesting that our method is likely to be less sensitive to high-dimension settings.

\section{Conclusion}
In this paper we have developed a computational strategy aiming to accelerate the IS estimation of a rare event probability, optimized with the CE method. At each level of the CE optimization process, a significant acceleration is obtained by selecting a surrogate from a  hierarchy of  score function approximations  according to the need for accuracy.  This need  is quantified using certified bounds on the approximation error. An asymptotic analysis proves that the proposed algorithm is guaranteed to converge in a worst-case scenario  at least as fast as state-of-the-art methods and that the reduced model selection is optimal in the sense it yields the minimal achievable estimator variance. In agreement with these theoretical arguments, numerical simulations quantify the gain brought by the proposed algorithm on a challenging pollutant transfer problem, where the high-fidelity score function is  modeled by a PDE and surrogates using RB.

\appendix

\section{A Sufficient Condition for Quantile Increase}\label{app:proofProp1}
~\vspace{-0.2cm}\\
 \begin{lemma}\label{rem:0} If  $\rho_{\bar \gamma}> \eta_{\bar \gamma,k_j}$, where $\rho_{\bar \gamma}=\langle   \unit_{\phi(\cdot) \ge \bar \gamma},  \nu_j \rangle$ and  $\eta_{\bar \gamma,k_j}$ is the probability  $\eta_{\bar \gamma,k_j}=\langle   \unit_{[ \gamma(  \nu_j,\rho_{\bar \gamma},\phi^{(k_j)}), \gamma(  \nu_j,\rho_{\bar \gamma},\phi^{(k_j)})+ \alpha_{k_j}( \nu_j)]},  \nu_j \rangle$,  
then there exists  $\rho>0$ such that  $\gamma(\nu_j, \rho, \phi^{(k_j)} )\ge \bar \gamma$ holds.\\ \vspace{-0.2cm}
 \end{lemma}
 
\proof{
if $\eta_{\bar \gamma,k_j} < \rho_{\bar \gamma}$, there exists $\rho \in (0 ,\rho_{\bar \gamma}- \eta_{\bar \gamma,k_j}] $ such that  \vspace{-0.15cm}
 \begin{align*}
\gamma( \nu_j,\rho,\phi^{(k_j)}) &= \max\{s\in \Rr : \langle  \unit_{\phi^{(k_j)}(\cdot)  < s} , \nu_j \rangle \le 1-\rho\},\\
&\ge \max\{s\in \Rr : \langle  \unit_{\phi^{(k_j)}(\cdot)  < s} , \nu_j \rangle \le 1-(\rho_{\bar \gamma}- \eta_{\bar \gamma,k_j})\},\\
&=  \max\{s\in \Rr : \langle  \unit_{\phi^{(k_j)}(\cdot)  < s- \alpha_{k_j}(\nu_j)} , \nu_j \rangle \le 1- \rho_{\bar \gamma}\},\\
&\ge \max\{s\in \Rr : \langle\unit_{\phi(\cdot)  < s} , \nu_j \rangle \le 1- \rho_{\bar \gamma}\}=\bar \gamma.
 \end{align*} 
The first equality is the quantile definition, the second equality follows from the definition of $\eta_{\bar \gamma,k_j}$ noticing that 
 $\gamma( \nu_j,\rho_{\bar \gamma},\phi^{(k_j)})+ \alpha_{k_j}(\nu_j)= \max\{s\in \Rr :  \langle \unit_{\phi^{(k_j)}(\cdot) < s- \alpha_{k_j}  },\nu_j \rangle \le 1-\rho_{\bar \gamma} \} ,$
 while the last one is obtained by making the change of variable $s'=s-\alpha_{k_j}$ in the definition of the quantile and using the definition of $ \rho_{\bar \gamma}$. The first inequality  is due to the fact that the expectation $\langle  \unit_{\phi^{(k_j)}(\cdot)  < s} , \nu_j \rangle$ is a non-decreasing function of $s$, while the last one is deduced from Remark~\ref{rem:1} by bounding the worst-case error. $\square$
}

\section{Inclusion Relations for the Relaxed Sets}\label{app:remarks}
We address two useful remarks. First, we notice that  the sequence of relaxed sets includes the sequence of original sets.    \\ \vspace{-0.2cm}
 \begin{remark}\label{cor:1}
 The definition of the sets implies the  inclusion:
 \begin{align}\label{eq:inclusion1}
A_{j}\subseteq  A^{(k_{j-1})}_{j}.
\end{align}


{
Indeed, 
Remark~\ref{rem:1} implies that $ \gamma(\nu_{j-1}, \rho_{j-1} ,\phi) \ge \gamma(\nu_{j-1}, \rho_{j-1} ,\phi^{(k_{j-1})})-\alpha_{k_{j-1}}(\nu_{j-1})$. 
If $x\in A_j$, then we obtain that 
$$
\phi^{(k)}(x)+\alpha_{k_{j-1}}(\nu_{j-1}) \ge \phi(x) \ge \gamma(\nu_{j-1}, \rho_{j-1} ,\phi) \ge\gamma(\nu_{j-1}, \rho_{j-1} ,\phi^{(k_{j-1})})-\alpha_{k_{j-1}}(\nu_{j-1}),
$$
which yields that $\phi^{(k_{j-1})}(x)\ge \gamma(\nu_{j-1}, \rho_{j-1} ,\phi^{(k_{j-1})})-2\alpha_{k_{j-1}}(\nu_{j-1})$, showing that $x \in A_j^{(k_j)}$ and therefore proving inclusion~\eqref{eq:inclusion1}. 
}
\\
\end{remark}


\noindent
We then note that  each element of  the sequence of relaxed sets includes the target~$A$. \\

\begin {remark}\label{rem:4}
By construction, we have $  A \subseteq   A_j$. Thus,  condition~\eqref{eq:inclusion1}
guarantees that   $  A \subseteq A^{(k_{j-1})}_j$. This condition is not only sufficient but also necessary  to guarantee  uniformly (over the  class of certified reduced-models) that $ A \subseteq  A^{(k_{j-1})}_j$, the latter condition being as already mentioned mandatory  to obtain a finite  variance of the IS estimator. Indeed, if  condition~\eqref{eq:inclusion1} does not hold, we can always build a reduced model with a maximum error $\alpha_{k_j}$ such that $  A\nsubseteq  A^{(k_{j-1})}_j   $ although  $  A \subseteq   A_j$. 
\end{remark}

 \section{Consistency of MC approximations}\label{app:prelim1}
    ~\\ \vspace{-0.2cm}
\begin{lemma}\label{prop:cvMC}
As $m \to \infty$, almost surely
\begin{itemize}
\item[{\it i)}]
 $\gamma(\hat \nu_j,\rho_j,\phi^{(k_j)})$ converges   to  $\gamma(\nu_j,\rho_j,\phi^{(k_j)})$,
\item[{\it ii)}] $ \alpha_{k_j}(\hat \nu_j)$ converges  to  $\alpha_{k_{j}}(\nu_{j})$,
\item[{\it iii)}] under Assumption B,~\eqref{eq:minDiv4_} converges  to~\eqref{eq:minDiv4_continue}.\\
\end{itemize} \vspace{-0.2cm}
  \end{lemma}

\proof{
Statement {\it i)} is shown in \cite{Homem-de-Mello02rareevent}. 
Statement  {\it ii)} 
 follows from Borel Cantelli Lemma~\cite{gut2006probability}. Indeed, denoting $M_{m_j}=\max(\epsilon_{k_j}(z_1),\ldots,\epsilon_{k_j}(z_{m_j}))$ where  $z_i$'s are i.i.d. samples of $\nu_j$ and denoting  $a= \max_{z\in \supp(\nu_j)}\epsilon_{k_j}(z)$, what needs to be shown is that  $\lim_{m_j \to \infty}M_{m_j}=a,$
with probability one. This happens to be the equivalent  of a zero probability of the event 
 $|M_{m_j}-a|>\tau\, \textrm{occurs infinitely often as } m_j \to \infty,$
 for any $\tau>0$. By Borel Cantelli Lemma, this holds if
 $ 
 \sum_{m_j=1}^\infty \langle \unit_{|M_{m_j}-a|>\tau}, \nu_j\rangle  < \infty.
  $
Moreover, independence of the $z_i$'s imply that $ \langle \unit_{|M_{m_j}-a|>\tau}, \nu_j\rangle =  \langle \unit_{M_{m_j}<a-\tau}, \nu_j\rangle = (\langle  \unit_{\epsilon_{k_j}(z_i)<a-\tau}, \nu_j\rangle)^{m_j}.$ We observe that by definition of $a$ , for any $\tau>0$,  we always have  $\langle  \unit_{\epsilon_{k_j}(z_i)<a-\tau}, \nu_j\rangle<1$.
Therefore, 
$
\sum_{m_j=1}^\infty \langle \unit_{|M_{m_j}-a|>\tau}, \nu_j\rangle=\frac{1}{1-\langle  \unit_{\epsilon_{k_j}(z_i)<a-\tau}, \nu_j\rangle}
$
 is a convergent geometric series  and we obtain the sought result. 
  Finally, to show   statement {\it iii)} we refer to the analysis used to establish the convergence of the standard CE method \cite[Proposition 5.3]{Homem-de-Mello02rareevent}. The latter proposition relies on the following statement:  under Assumption B, assume a set $\hat C \subseteq \Rr^p$ converges almost surely as $m\to \infty$  towards  $C \subseteq \Rr^p$; then  almost surely we have that \vspace{-0.2cm}
$$
  \lim_{m\to \infty}\argmax_{\nu^\theta \in \mathcal{V}} \frac{1}{m} \sum_{i=1}^m \unit_{\hat C(z_i)} \frac{\mu(z_i)}{\nu_{j-1}(z_i)} \ln \nu^\theta(z_i)
=\argmax_{\nu^\theta \in \mathcal{V}}\langle  \unit_C \frac{\mu}{\nu_{j-1}} \ln \nu^\theta , \nu_{j-1}\rangle, \vspace{-0.2cm}$$
where $z_i$'s are {\it i.i.d.} samples drawn according to $\nu_{j-1}$. 
Setting $\hat C=\hat A^{(k_{j-1})}_{j}$ and $ C= A^{(k_{j-1})}_{j}$ and using the identity \vspace{-0.2cm}
$$ \argmax_{\nu^\theta \in \mathcal{V}}\langle  \unit_{ A^{(k_{j-1})}_j} \frac{\mu}{\nu_{j-1}} \ln \nu^\theta , \nu_{j-1}\rangle=\argmax_{\nu^\theta \in \mathcal{V}}\langle  \unit_{ A^{(k_{j-1})}_j}  \ln \nu^\theta , \mu\rangle, \vspace{-0.2cm}$$
we see that, to prove that approximation~\eqref{eq:minDiv4_} converges  almost surely towards the optimal solution~\eqref{eq:minDiv4_continue},  it is  sufficient to show that the set $  \hat A^{(k_{j-1})}_{j}$  converges almost surely towards $  A^{(k_{j-1})}_{j}$. This is equivalent to show that 
$  \gamma( \hat \nu_{j-1},\rho_{j-1},\phi^{(k_{j-1})}) $ and $ \alpha_{k_{j-1}}(\hat \nu_{j-1})$ converge almost surely  towards $  \gamma( \nu_{j-1},\rho_{j-1},\phi^{(k_{j-1})}) $ and $ \alpha_{k_{j-1}}(\nu_{j-1})$, which is precisely statement {\it i)} and {\it ii)}. $\square$

  }
  

\section{Proof of the Second Statement of  Theorem~\ref{theorem:0}}\label{app:0}
{

Assuming that $\nu_{\hat A_j}^{\star} \in \mathcal{V}$ and $m_j$ large enough, the  variance  writes  \vspace{-0.1cm}
\begin{align*}
\mathbb{E}[(  p^{IS}_{A,\nu_j}-p_A)^2]&=\frac{1}{m_j}\textrm{var}( \unit_{A}\frac{\mu}{\nu_j}, \nu_j)  = \frac{1}{m_j}\left( \langle  (\unit_{A}\frac{\mu}{\nu_j})^2, \nu_j\rangle -p_A^2\right)\\
&=\frac{1}{m_j}\left(  \langle \frac{\unit_{A}\mu}{\nu_j} ,\mu\rangle  -p_A^2\right)=\frac{1}{m_j}\left(   p_{\hat A_j}\langle \frac{\unit_{A}}{\unit_{\hat A_j}} ,\mu \rangle  -p_A^2  \right)
=\frac{p_A}{m_j}\left(   p_{\hat A_j}  -p_A  \right),
\end{align*} 
where the second equality follows from the fact that estimator~\eqref{eq:IS} is unbiased, the fourth equality substitutes $\nu_j$ by the zero-variance density~\eqref{eq:optDensity} because of the assumption $\nu_{\hat A_j}^{\star} \in \mathcal{V}$ and the asymptotic hypothesis, and the fifth equality is obtained using the condition  $\hat A_j \supseteq   A$  fulfilled by the $\hat A_j$'s.\\\vspace{-0.cm}

}

\textbf{Acknowledgements}\\
The author wishes to thank Fr\'ed\'eric C\'erou and Mathias Rousset for fruit-full discussions
on Monte-Carlo approximations and rare event simulation. He also thanks anonymous referees for their insightful comments which  helped in improving the paper. Finally, he is sincerely grateful to B.L. for providing him a great deal of understanding. \vspace{-0.2cm}

\bibliographystyle{IEEEbib}
\bibliography{./bibtex}

\end{document}